# Novel active micromixer platform based on Lorentz force for lab-on-a-chip application

Aniket Kandalkar[a], Nachiket Pathak[a,b], Atharva Kulkarni[c], Amit Morarka*[c]
[a]Institute of Bioinformatics and Biotechnology, Savitribai Phule Pune University, Ganeshkhind, Pune, Maharashtra, INDIA - 411007
[b]Indian Institute of Science, C.V. Raman Avenue, Bengaluru, Karnataka, INDIA-560012
[c]Department of Electronics and Instrumentation Science, Savitribai Phule Pune University, Ganeshkhind, Pune, Maharashtra, INDIA- 411007
*amitmorarka@gmail.com

## Abstract

Mixing in an active micromixer was achieved using Lorentz force-assisted actuation of an enameled copper wire. A single-step template-assisted soft lithography technique was used to construct the mixing chamber. The chamber had a volume of 1.86µl. Application of a Square wave alternating current in tandem with tension in the wire provided the necessary conditions for the resonant oscillation frequency of the wire. The repeatability of the ratio of higher harmonics to the fundamental frequencies of the oscillating wire conforms to standardization of the device fabrication, assembly, and functionality. Simulations and experiments were performed to validate uniform temperature distribution in the mixing chamber. Real-time optical detection of the sample assisted in sensing the completion of chemical reactions in the chamber. Mixing of various aqueous based chemical reactions was performed. It was found that mixing efficiency was greater than 95 percent. Multiple devices were fabricated to show the usability and reproducibility of the system.

## 1. Introduction

Microfluidics is an approach to control fluid dynamics and its interaction at a scale of $10^{-6}$ liters and below. At this scale, the surface and viscous forces dominate over the inertial forces in the fluid. This uniqueness of the microfluidic systems has been intensively studied and ubiquitously used in various disciplines of sciences Ref. 1,2. Advances are being made to reduce the large-scale volume of a laboratory reaction into a microfluidic system. This would equally help to perform complex reactions Ref. 3–5. There are many chemical reactions which demand small volumes as the reagents may be valuable, or their reaction product yield is less. Hence the microfluidic system is a suitable system to study such reactions Ref. 6,7.

It is very well established Ref. 8 that in microfluidic systems, fluids have a low Reynolds number and exhibit laminar flow conditions. Therefore, the mixing of fluids in the microfluidic chamber becomes a challenging task as there is no turbulent flow. To accelerate the mixing in the system, active and passive microfluidic mixers have been developed. In a passive mixer, there are no moving

*Abbreviations*: PDMS, Polydimethylsiloxane; MC, main chamber; D80, enameled copper wire of diameter 80µm; LED, light emitting diode; PT-100, resistance temperature detectors; NaOH, sodium hydroxide; HCl, hydrochloric acid; $KMnO_4$, potassium permanganate.

components while the internal architecture of the device is such that it generates turbulence as the fluid flows through the channel. Whereas, in an active mixer there are actuators or moving elements that generate turbulent flow hence accelerating mixing.

A number of reports demonstrate active microfluidic devices having a magnetic stirrer in the microfluidic chamber Ref. 9–11. In another study Ref. 12 ferrous nanoparticles were introduced in the sample of interest. The stirrers and particles moved in response to the change in direction of an external magnetic field. Their motion could efficiently enable mixing in the system. Additionally, microfluidic systems using ferrofluids have also been demonstrated, which are actuated under changing electromagnetic fields Ref. 13. Similarly, Lorentz force was exploited to perform the trapping of cells and mixing in the microfluidic system Ref. 14,15. There are reports on the fabrication of a microfluidic system using the conventional photolithography technique on Polydimethylsiloxane (PDMS**)** and Silicon nitride Ref. 16,17. Their device had a microplate embedded inside the PDMS microchannel. The motion of the plate was due to Lorentz force. The operation of this device required a 140 kHz signal for its oscillatory motion. Its movement generated vortices which helped to trap Human embryonic kidney cells and could also show micromixing in the chamber.

In microfluidic devices, it is always observed that precise control over the physical parameters eases the device operation as well as good control over the results. Additionally, the device fabrication should be cost-effective, simple, and made from readily available materials. Taking all these factors into consideration, a current carrying wire in a magnetic field is the simplest system to observe Lorentz Force. This could be utilized as an active mixing method in a microfluidic device made from template-assisted soft lithography technique. "Lorentz Force" is mathematically represented by Eq. (1) Ref. 18

$$\vec{F} = \mathrm{I}\,(\vec{\mathrm{l}} \times \vec{\mathrm{B}}). \qquad (1)$$

Based on this, a novel method was conceptualized to fabricate a cost-effective active micromixer using only template-assisted microchannel fabrication. The mixing could be done using Lorentz force acting on a current-carrying conductor placed inside the microchannel under a known tension. To the best of our knowledge till date, such a device does not exist.

In this study, a micromixer was constructed using template-assisted soft lithography Ref. 19 and was characterized. An enameled copper wire passing through the chamber exposed to an external magnetic field, executed oscillatory motion upon excitation by a square wave AC signal. Its motion caused mixing in the microchamber. The simple assembly process of the device makes it easily reusable. As depicted in Fig. 1, the Lorentz force was responsible for the oscillatory motion of wire in the chamber with an oscillation frequency less than 1 kHz. The volume of the main chamber (MC) is 1.86µl. Along with active mixing, optical sensing and temperature regulation and sensing were achieved. Therefore, this device could be used for Lab-on-a-chip applications.

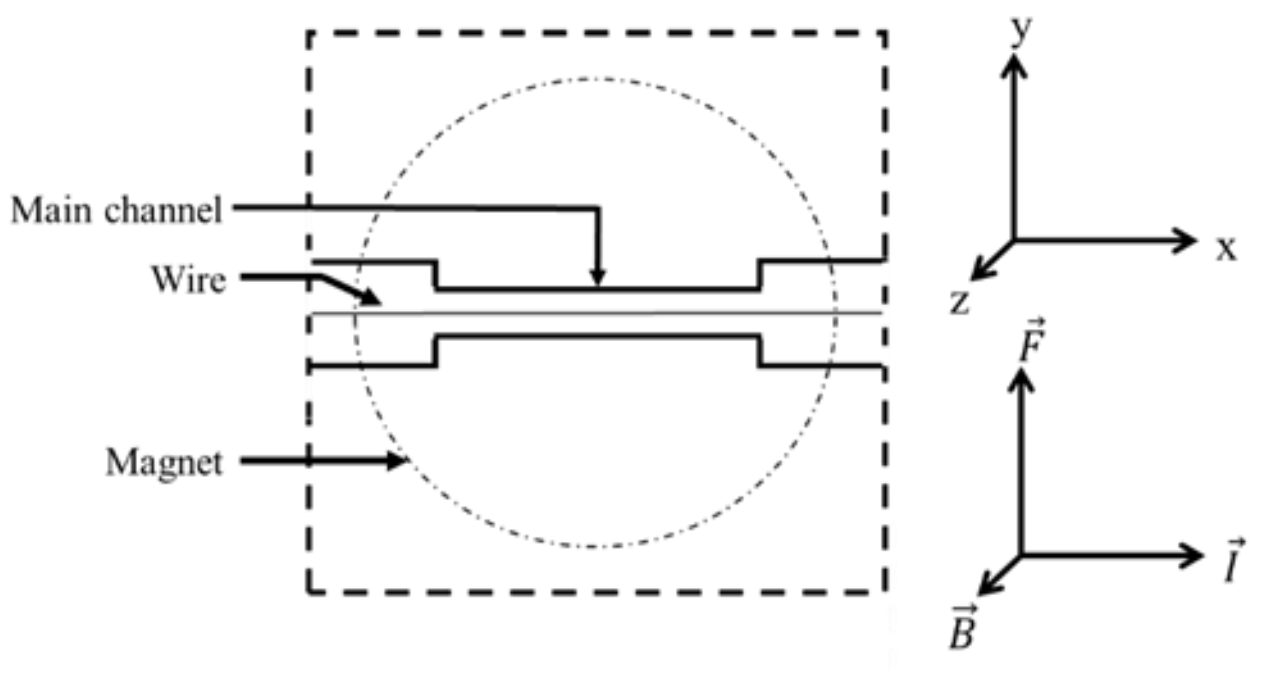

**Fig.1**. Schematic of device representing direction of magnetic field which is perpendicular to the plane of the paper (z-axis) and the direction of the square wave AC signal along the length of wire (in the main channel) (x-axis). This imparts a resultant Lorentz force along y-axis which lies in the plane of the paper. Square made of dashed line represents the entire PDMS device.

## 2. Experimental

The device was fabricated by template-assisted soft lithography Ref. 19 using PDMS polymer. The casting mold was made from off the shelf materials. The fabricated device had a cylindrical cavity of diameter 450µm and length 11.7mm at the center. This is the MC in which the mixing was achieved. The MC was continuous with the main channel which was passing through the device as shown in Fig. 2. An enameled copper wire of diameter 80µm (D80) and length 195mm was positioned co-axially which went through the main channel. The two terminals of the wire were rigidly fixed to maintain mechanical tension and electrical conductivity in the wire. The main channel provided support to mount the bushings which ensured that the wire closely coincided with the axis of the channel. The main channel was sealed at the ends using grease to avoid leaks.

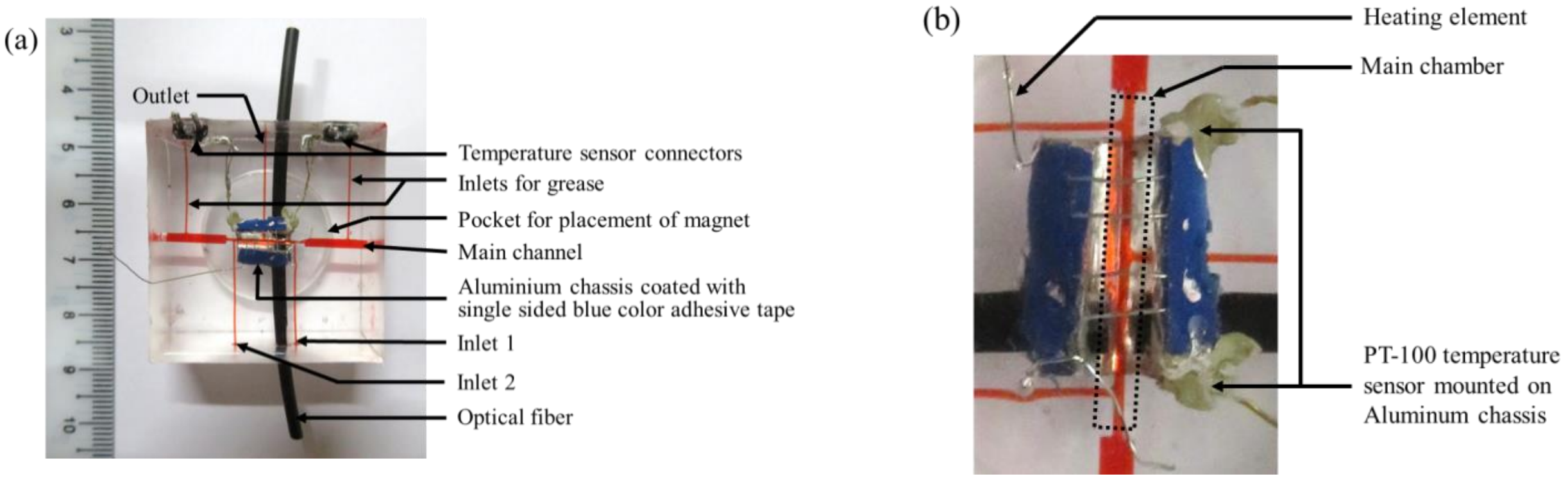

**Fig. 2**. Images of (a)Final microfluidic device with channels filled with red ink for its visibility. (b)Inset of Main chamber (marked by dotted box) having a volume of 1.86µl**.** A detailed explanation of the device is provided in supplementary section 1.

A platform was designed and fabricated on which the device was mounted firmly to carry out the study. Fig. 3 shows the mounting scheme of the device with wire under tension. The tension in the wire was regulated by the electromagnet which pulled a ferromagnetic weight of 2.5g attached to it. For the real-time monitoring of the tension, a load cell connected to HX711 ADC module was used along with Arduino UNO to read the tension values and was calibrated. A cylindrical pocket was made at the bottom of the device to place a rare earth magnet 4mm below the MC. The measured magnetic field was 1960 Gauss at this distance (Digital Gaussmeter DGM-102, Scientific equipment Roorkee). A square wave AC electrical signal from Agilent 33220A 20MHz function generator was used to actuate the wire. The signal was buffered through H-bridge IC L293D as shown in the circuit diagram in Fig. 3. This configuration of wire and magnet provided the required Lorentz force which was responsible for the oscillatory motion of the wire which caused the mixing of liquids in the MC.

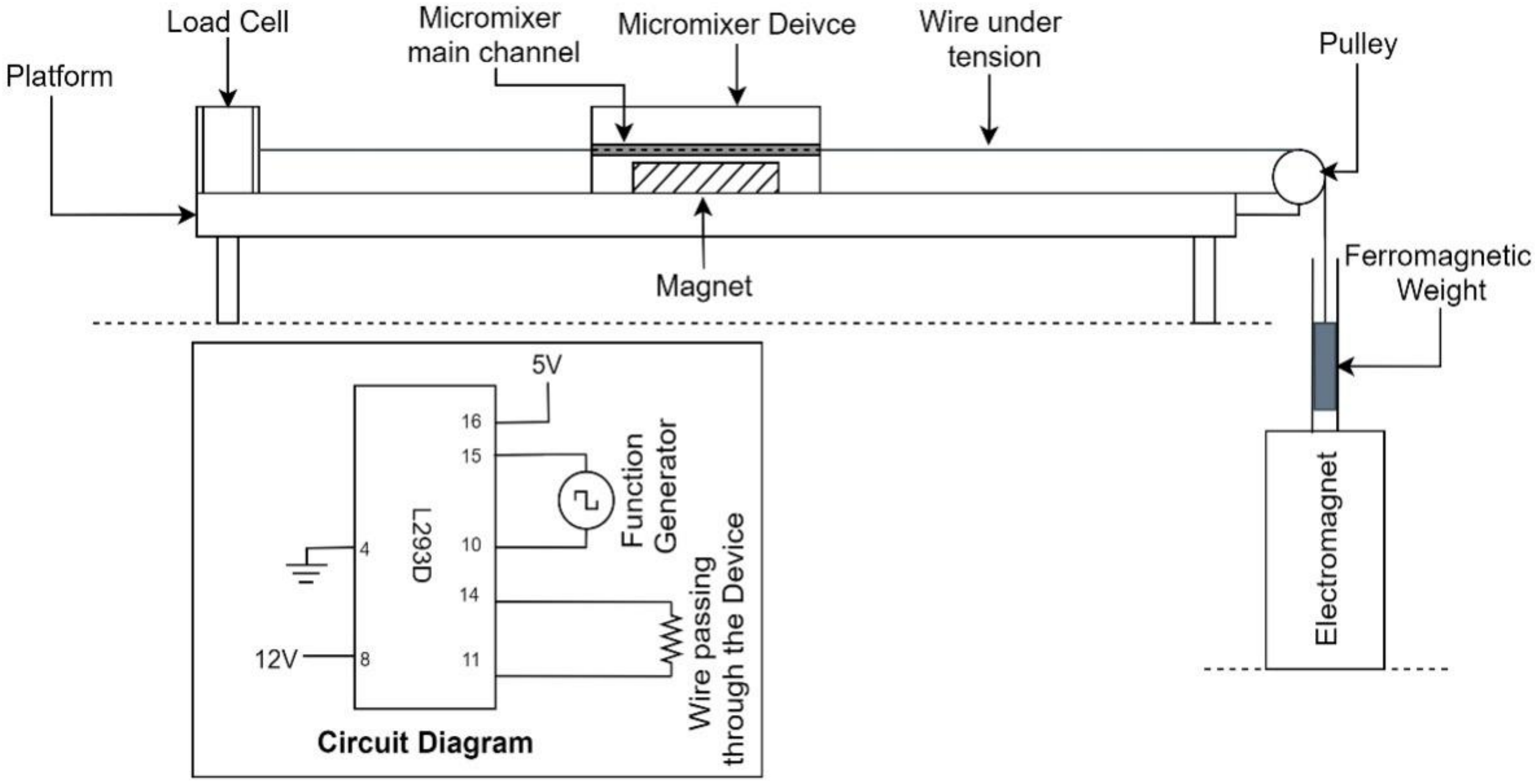

**Fig. 3**. Side view schematic of device mounted on the platform and the circuit diagram.

The characterization of the frequency of the oscillations of wire was done to find the resonant frequency for different applied tensions. The wire tension and the frequency were studied over a range. Three discrete values for tension 0.0392N, 0.0686N, 0.1078N were considered, and the frequency was manually swept from 1Hz to 1 kHz. This was done using acoustic and optical stroboscopic recordings of the oscillating wire.

For regulating the temperature inside the MC, Ohmic heating was implemented. This was done by winding four turns of the steel wire (diameter 150µm) around the aluminium chassis. The chassis had inner and outer diameters of 3mm and 6.33mm respectively and a length of 9mm. Single side adhesive tape coated over aluminium chassis provided the required electrical insulation between the steel wire and the aluminium chassis. Electrical power was applied to the steel wire (heating element). The temperature was monitored by two CRZ2005R-100 Resistance Temperature Detectors (PT-100), Hayashi Denko Co. Ltd. attached at the two axial ends of the chassis. Optical sensing in the device was done using an Ocean Optic CCD HR4000 spectrophotometer and Light-emitting diode (LED) (Peak wavelength 538nm) as the source. The light was coupled through the MC using optical fiber keeping 450µm path length. For details of fabrication and assembly of the micromixing device refer to supplementary section 1.

Three case studies were done for validating the application of the device. In all the studies amplitude of the wire oscillation was set to maximum. Two solutions in a 1:1 volumetric ratio were pumped manually inside the MC using a Hamilton syringe through the inlets of the device.

## 3. Results and discussion

In this work, we fabricated a novel and simple active microfluidic mixer device by utilizing a single-step soft lithography technique. An oscillating copper wire (D80) performed mixing in the MC as it experienced maximum Lorentz force due to the unique architecture of the micromixer. The oscillations in the wire were generated by the changing direction of the Lorentz force with the frequency of the square wave alternating current riding on the wire. The oscillations therefore obtained were parallel to the plane of the platform on which the device was mounted.

A current carrying wire (D80) is expected to cause Ohmic heating, therefore heating the MC. As the temperature measurement of the straight wire was not directly possible, a simulation was performed in COMSOL (supplementary section 2). The simulation included the complete device with MC having length of 11.7mm and the D80 passing through it. The results show that the temperature of the D80 increased to 28.2°C and that of MC increased by an average of 1°C from the room temperature (Fig. 4). Since the rise in temperature is small the end results of mixing are not affected significantly. Taking into consideration the temperature rise of MC due to current flow in D80, the amplitude of wire oscillation, and MC volume, the current and length of the D80 was fixed to 0.2A and 10mm respectively.

To achieve efficient mixing, it is required that the wire oscillations should be at resonance. Hence, experiments were performed to optimize the amplitude, frequency, and tension of the oscillating wire to ensure that the system operates at its resonant frequency.

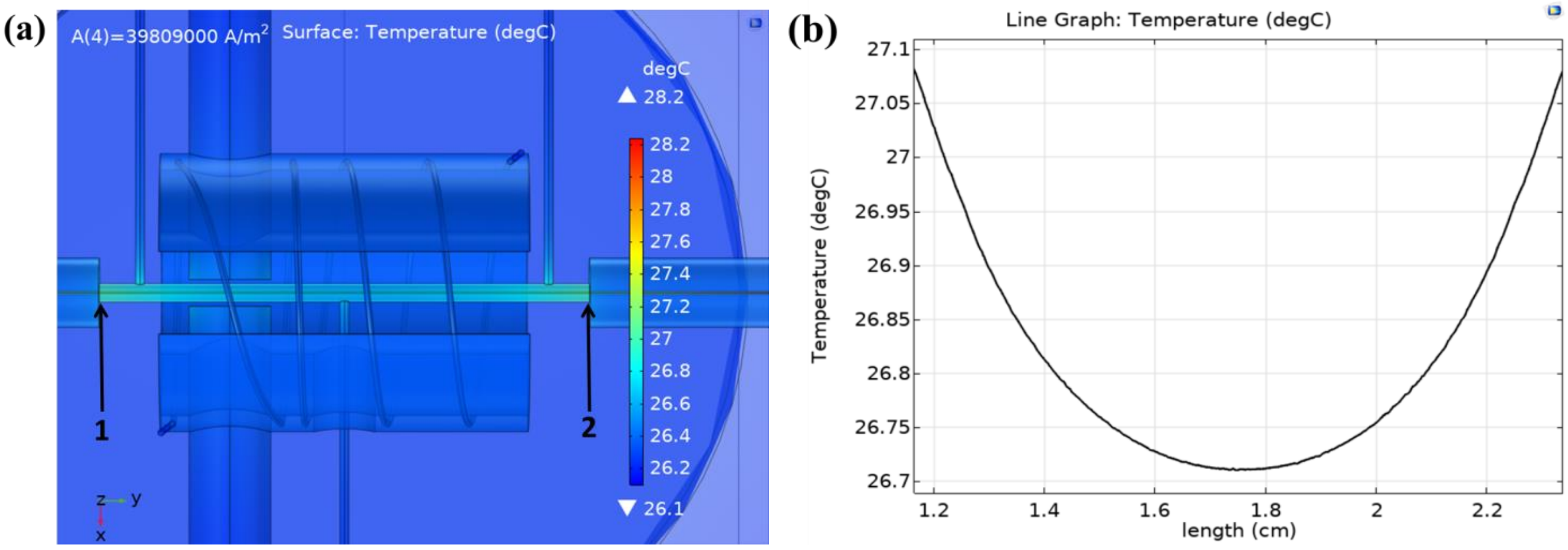

**Fig.4**. (a) Thermal plot from the simulation result of heating of MC caused by the D80 wire. The channel between the points 1, 2 is the MC. (b) Line graph of temperature distribution along the complete length of D80 wire inside MC.

The expression governing the dynamics of the oscillating wire under tension in a dissipative environment could be given by Eq. (2). Ref. 20

$$\left(\rho A + C_a \rho_l A\right) \frac{d^2 \omega}{dt^2} - F_x \frac{\partial^2 \omega}{\partial x^2} = f(\omega, t) - F_{drag}, \tag{2}$$

Where the displacement of the wire is $\omega$, $F_x$ is the tension in the wire, $\rho A$ is the mass per unit length of the wire, $C_a \rho_l A$ takes into consideration the added mass per unit length due to the presence of fluid, $F_{drag}$ is the drag force due to damping in the fluid, $f(\omega, t)$ is the driving force. In the present work $f(\omega, t)$ is the Lorentz force as described in equation (1) which is responsible for the amplitude of the wire oscillation, controlled by the current passing through the wire. This same current is also responsible to cause a 2°C rise in temperature of the D80. With this temperature change, there is no significant change in its mass per unit length Ref. 21. Therefore, the amplitude and resonance frequency of the wire remains unaffected.

The D80 wire had a mass per unit length of $46.91 \times 10^{-6}$ Kg/m. It was empirically found that removing sag in the wire having a length of 195mm required a minimum tension of 0.0245N. With the continuation of the experiment, three values of tension 0.0392N, 0.0686N, and 0.1078N were selected for which resonant frequencies along with harmonics were recorded. Resonating frequencies were studied using synchronized acoustic and optical stroboscopic techniques (supplementary section 3). Fourier analysis was performed on the acoustic data of the wire oscillations using MATLAB.

**Table I**
Amplitude of oscillating wire at various frequencies.

| Tension (N) | Harmonics | Device 1 | | | Device 2 | | | Device 3 | | |
|---|---|---|---|---|---|---|---|---|---|---|
| | | Frequency (Hz) | Amplitude (μm) | Ratio [a] | Frequency (Hz) | Amplitude (μm) | Ratio [a] | Frequency (Hz) | Amplitude (μm) | Ratio [a] |
| 0.0392 | Fundamental | 222 | 58 | - | 206 | 20 | - | 229 | 39 | - |
| | 2nd harmonic | 587 | 87 | 2.64 | 525 | 45 | 2.54 | 606 | 43 | 2.64 |
| | 3rd harmonic | 674 | 187 | 3.03 | 611 | 168 | 2.96 | 688 | 107 | 3.00 |
| 0.0686 | Fundamental | 267 | 58 | - | 274 | 59 | - | 270 | 21 | - |
| | 2nd harmonic | 797 | 141 | 2.98 | 839 | 78 | 3.06 | 810 | 64 | 3.00 |
| 0.1078 | Fundamental | 466 | 42 | - | 320 | 59 | - | 337 | 21 | - |
| | 2nd harmonic | 994 | 152 | 2.13 | 976 | 98 | 3.05 | 990 | 86 | 2.93 |

[a] Ratio of harmonics to the fundamental frequency

Comparing the results obtained from acoustic and video recording techniques provided the resonating frequencies for the applied tension. This analysis also shows that the wire oscillation frequency was the same as that of the input square wave signal frequency. It was found that the wire oscillations were maximum at a tension of 0.0392N for a frequency of 658Hz which is the average of three devices that were constructed (Table I). Interestingly, all the devices showed an average value in the range of 2.7-3.0 which is the ratio of higher harmonics to the fundamental frequencies of the oscillating wire. This shows that the devices are in unison with respect to the fabrication process, assembly, and working. This could conform to standardization as shown in Table S3 of supplementary material. Further in all the case studies, these two parameters were maintained.

For the system to stand for Lab-on-a-Chip application thermal regulation and optical sensing are must. Therefore, for heating the MC and uniform dissipation of the heat in the chamber, we proposed to integrate a hollow aluminium chassis into the device. To confirm this, thermal simulations using COMSOL were done. The geometry was constructed similar to the device that was designed along with the proposed aluminium chassis. The heating element was the source and appropriate power was offered in units of $W/m^2$. The stationary study of the system showed that there is uniform temperature distribution with a certain amount of temperature gradient between the chassis and the MC. The results of the same are presented in Fig. 5. The results suggested that the aluminium chassis provides a uniform temperature source surrounding the MC. The temperature distribution in the MC is uniform as shown in Fig.5(c). Hence, a hollow aluminium chassis was integrated into the device.

It will be naive not to consider the presence of a thermal gradient in between the chassis and the MC in the fabricated device. To facilitate the temperature measurement of the chassis, a pair of PT-100 sensors was attached to it. The device thermal calibration was done by simultaneously monitoring the temperature from these two PT-100 and a thermocouple which was inserted in the MC. The thermocouple was constructed in-house and calibrated (figure S11 in supplementary section 4). The thermocouple had a diameter comparable to the MC diameter. Since PDMS is highly elastic, the inserted thermocouple did not cause any permanent deformation to it. For the micromixer device, the range of the thermal calibration was from 24°C to 73°C. These calibrations were referred to while performing temperature-dependent chemical reactions (figure S12 in supplementary material).

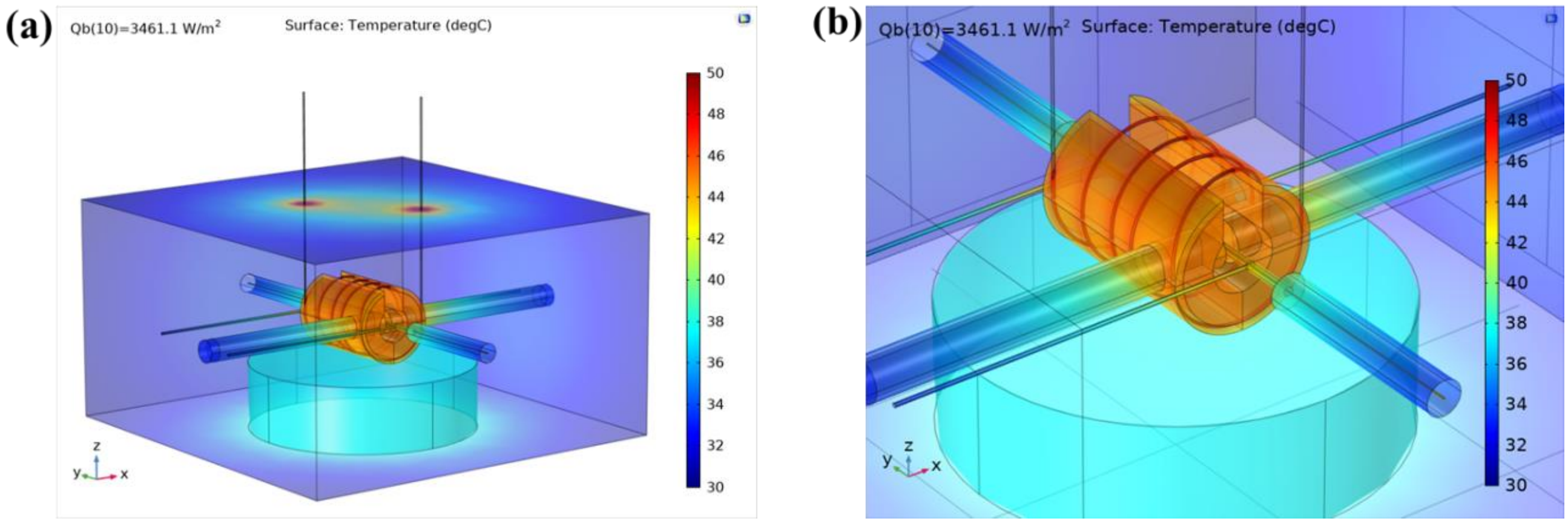

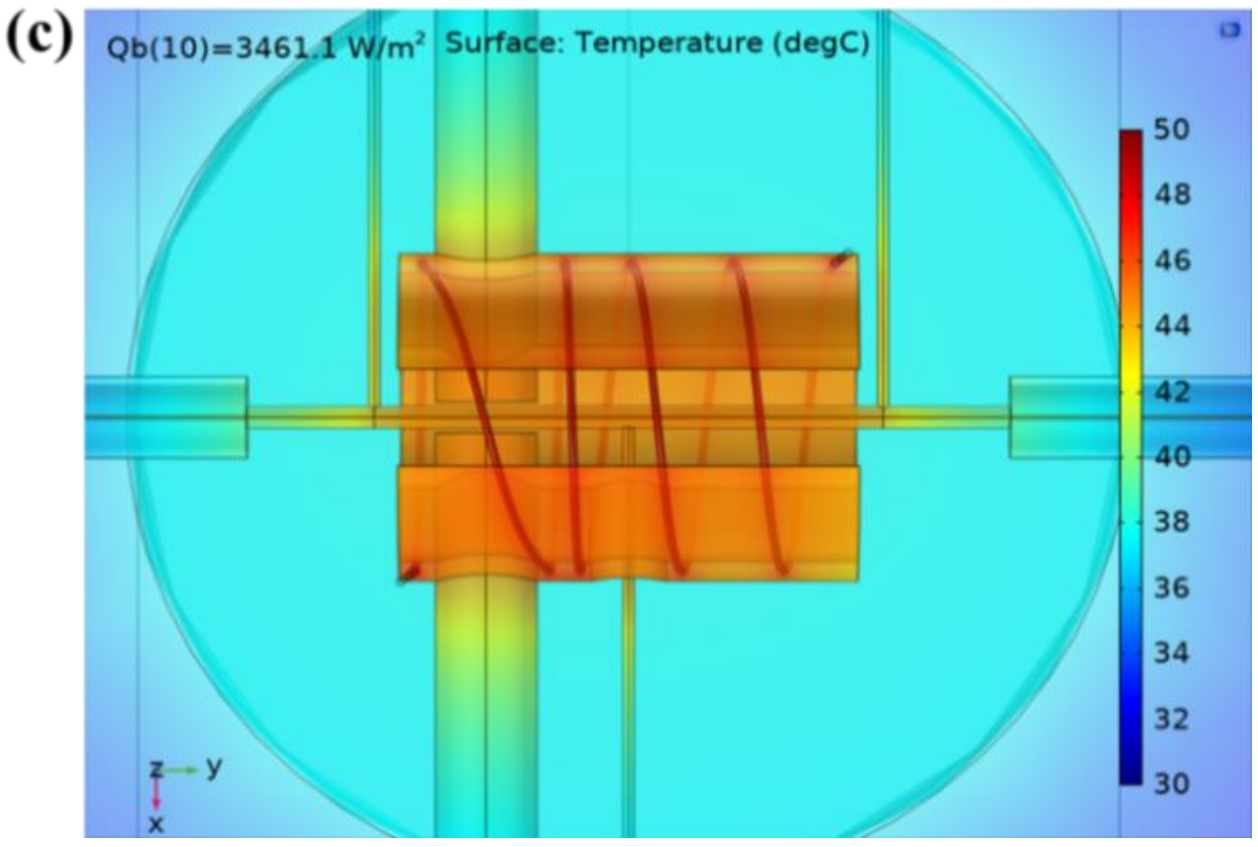

**Fig. 5**. Thermal simulation of the device to understand the temperature distribution. The surface temperature plots of the MC were simulated for 45°C as a reaction was performed at this temperature further in the study. (a) Surface temperature (°C) plot of complete device (perspective view). (b) Enlarged view of aluminium chassis (perspective view). (c) Top view of aluminium chassis with main channel visible. Additional analysis is provided in figure S10 of supplementary material

The optical setup was made to measure the completion of the mixing and reactions in the MC. LED was used as a source which was coupled to the system using the optical fiber. Optical calibration of three different devices was done to check for the repeatability of the optical assembly. The complete results of the same are shown in supplementary section 5. The deviation in the data acquired from the three devices must be due to the variation in the coupling of the optical fiber as it was done manually.

All the previously mentioned types of calibration results correlate with all the devices that were characterized. To show the usability and repeatability of the device, three case studies were performed using chemical reactions in three different devices.

### 3.1. Case 1: Mixing of ink and water

As described in the experimental section, mixing of 10% red ink from the stock solution and distilled water was done in a device. The interfaces of the solutions were brought in contact with each other by manually pumping solutions in the device. The start time of the reaction was considered when the interfaces touched each other. The sample was left undisturbed for it to mix by diffusion and optical sensing was done simultaneously. It was used as a measure to determine the completion of the reaction by measuring the intensity counts of the incident light passing through the sample. Since the optical assembly was off-centered, two different experiments were performed. In one experiment water was pumped on the side of the optical sensor and in another experiment by injecting ink on the side of the sensor. This was done to show that solutions were uniformly mixed by the wire and the mixing in the MC was not dependent on which side the samples were injected. Fig.6 shows the results of the two experiments. Solutions mixed in a 1:1 volume ratio outside the chamber was the control.

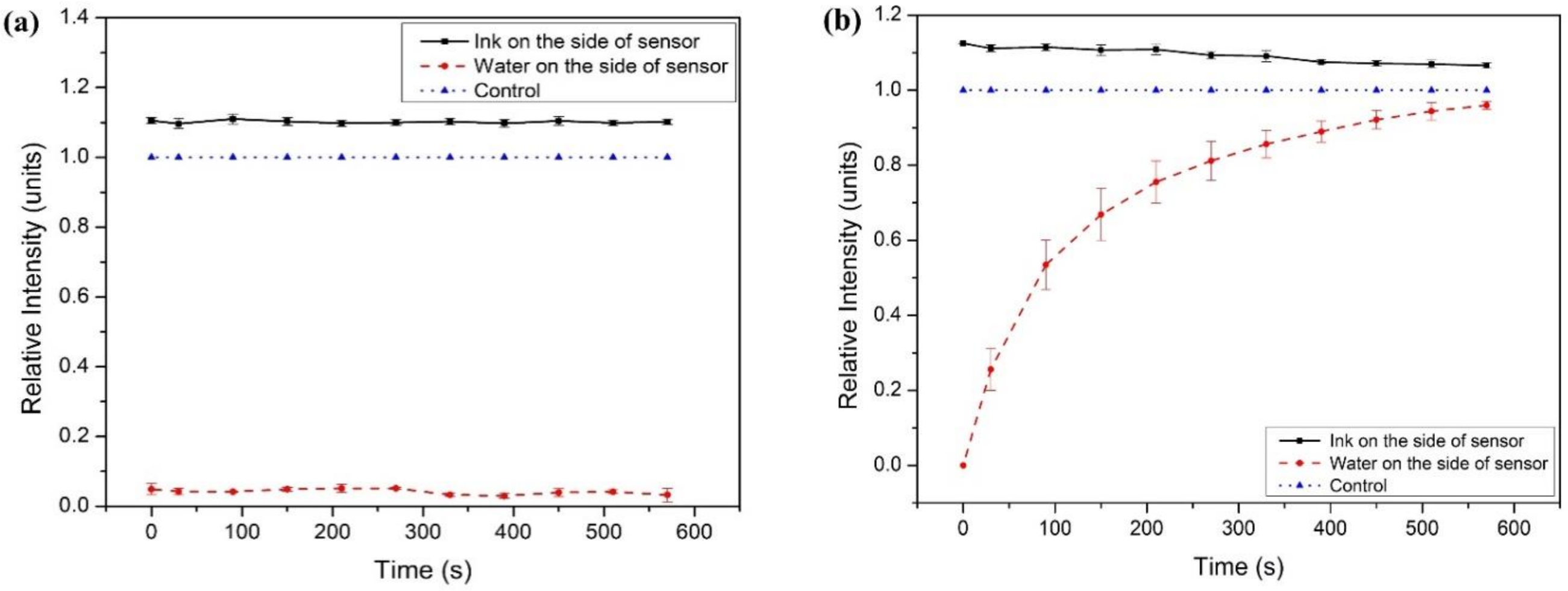

**Fig. 6**. Ink and water mixing in the device. (a) The solutions were allowed to mix by diffusion. (b) The active mixing was done by the oscillating wire. In the graph relative intensity count of 1.0 represents solutions are completely mixed (control) while relative intensity count of 0.0 represents completely transparent solution. Any value higher than 1 represents concentrated solution than the control. The reaction was repeated thrice in the same device and the average values were plotted in graphs.

The acquired data was compared with the control experiment by measuring the relative intensity counts in the device. The mixing caused by diffusion in both experiments was incomplete even after 600 seconds whereas active mixing achieved an average of 95% of the control solution in the same time. Images of the reaction before and after completion are shown in supplementary section 6. This helped to prove that there is no concentration gradient along the length of the MC after it was actively mixed.

### 3.2. Case 2: Acid-base reaction

As described in the experimental section, the experiment was conducted in another device using an acid-base system where phenolphthalein was used as an indicator. Phenolphthalein was already mixed with Sodium hydroxide (NaOH) having a concentration of 0.024M, giving it a pink color.

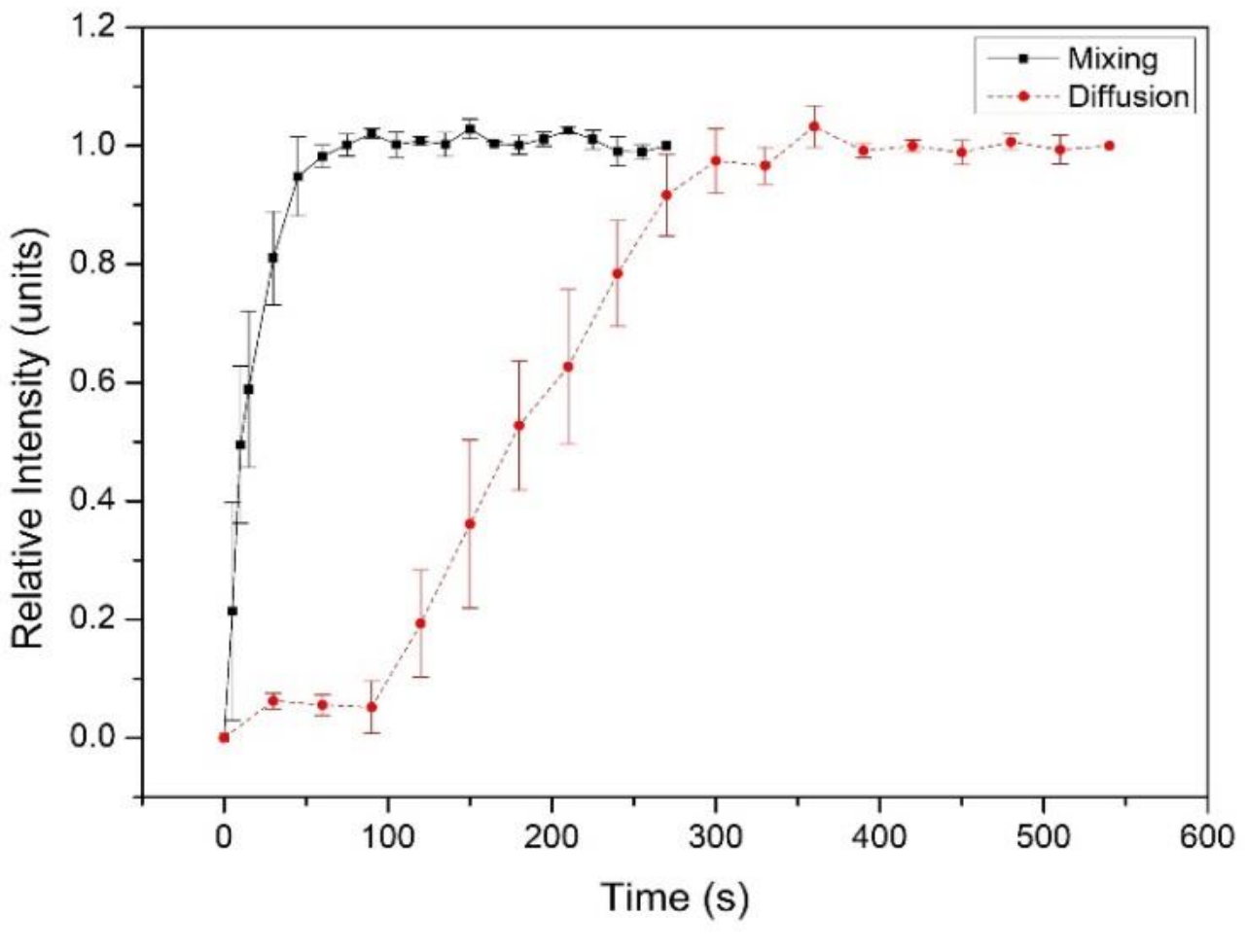

**Fig 7**. Mixing of acid and base in the Device. The reaction was repeated thrice in the same device and the average values were plotted in graph.

.

Hydrochloric acid (HCl) of 0.4M was used. In this case, NaOH was pumped through the side of the sensor. The product of this reaction was a colorless solution due to the change in pH, making it acidic. From the data as seen in Fig.7, it can be observed that in both the cases where diffusion and active mixing is taking place the resultant product is a colorless solution having an average relative intensity count of 1. In the case of active mixing, the color change is very fast as compared to diffusion. The solution becomes colorless in the first 50s in case of active mixing whereas in the case of diffusion it took up to 300s. This implies that in this case, the active mixing is 6 times faster than the diffusion-assisted mixing.

### 3.3. Case 3: Temperature-dependent reaction of Potassium Permanganate and Oxalic Acid

As described in the experimental section, the third reaction was carried out in a device where the temperature of the MC was regulated as discussed earlier. It has been reported that oxidation of oxalic acid in presence of potassium permanganate ($KMnO_4$) is a temperature-dependent reaction Ref. 22. The reaction rate increases with the rise in temperature and the solutions turns transparent from the pink colored solution. The temperature was maintained in the MC throughout the reaction. Along with active mixing for a fixed time of 570s, the light intensity counts were monitored. In Fig.8, relative light intensity counts are reported with respect to time. Images of the reaction before and after completion are shown in supplementary section 6. It was observed that at 45°C the reaction rate increased as compared to the reaction that occurred at 20°C. The completion of the reaction at a higher temperature was achieved after the lapse of 300s, and it took more than 570s for the reaction to complete at 20°C.

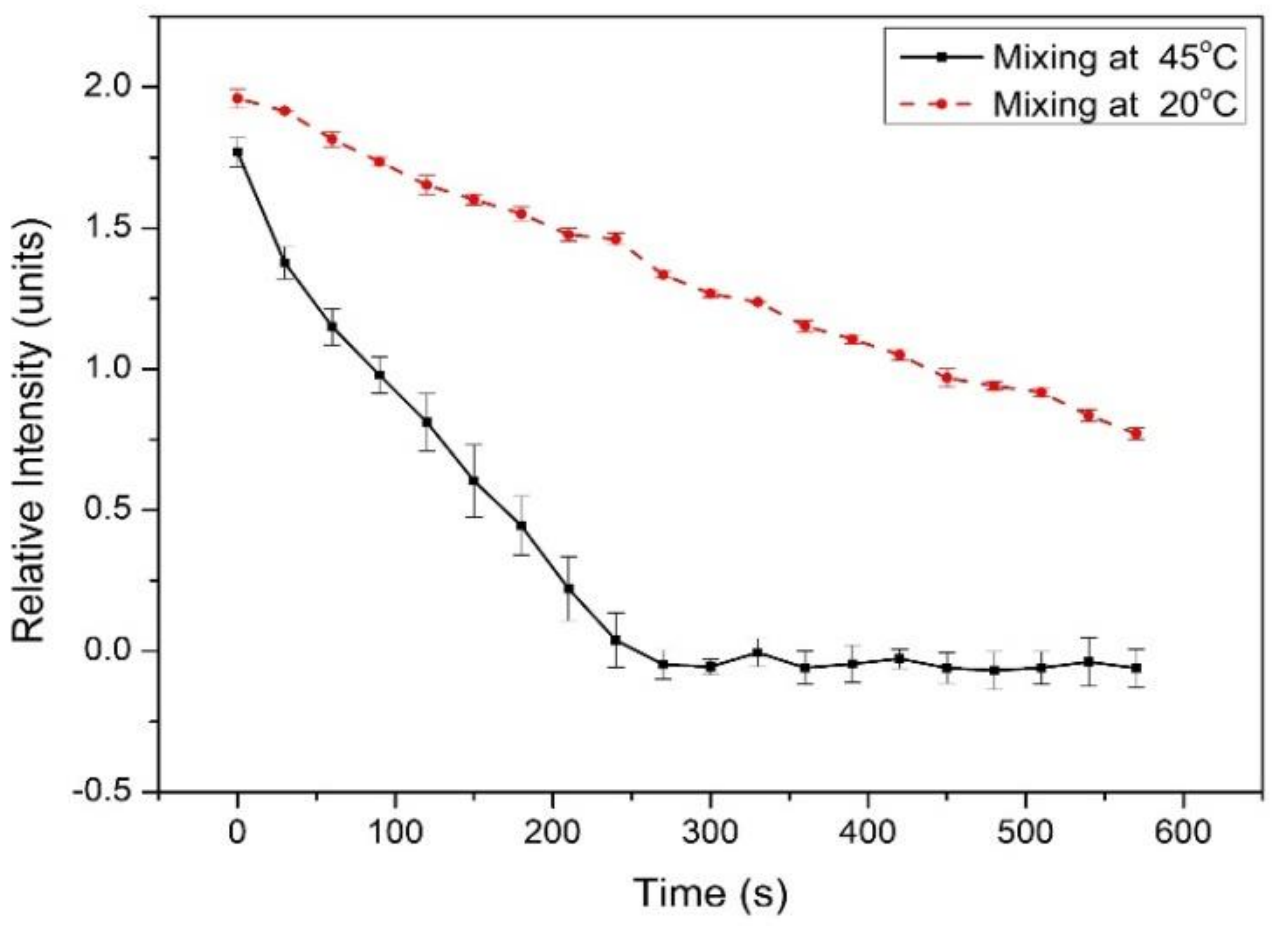

**Fig 8**. Temperature dependent reaction between oxalic acid (0.017M) and potassium permanganate (0.002M). Here relative intensity count of 0 represents colorless solution and relative light intensity count of 2 represent $KMnO_4$ having concentration of 0.002M. The relative intensity of count 0.0 represents completely mixed solution which was transparent. The reaction was repeated thrice in the same device and the average values were plotted in graph.

A comparative study of the existing micromixers is presented in Table II. After a detailed study, it was found that a very small number of micromixer devices exist which are based on the Lorentz force. Their constructed prototypes use conventional fabrication techniques like Silicon micromachining using photolithography, Low Temperature Co-fired ceramic and Template assisted soft lithography. This makes the fabrication process expensive, tedious and time consuming. The fabrication of this device is a single-step procedure, using only template-assisted soft lithography. Also, none of the reported devices claim to have integrated sensing mechanisms along with micromixing, as it is always preferred for a Lab-on-a-Chip system to carry out sensing for its applications.

**Table II**
Comparative analysis of Lorentz force based micromixer.

| Sr. No. | Microchannel/ Microchamber dimensions (μm) | Fabrication Process | Lowest Background fluid Flow velocity | Sensing | Functionality | References |
|---|---|---|---|---|---|---|
| 1 | 50 × 1,000 × 1,000 (height × length × width) | Soft Lithography, Multistep Conventional photolithography | 0.078 mm/s | None | Micromixing, Cell trapping | Ref. 16,17 |
| 2 | 10,000 × 10,000 × not mentioned (height × length × width) | Multistep Conventional photolithography | Not mentioned | None | Micromixing | Ref. 15 |
| 3 | 2,000 × 85,000 × 4,000 (height × length × width) | CNC Milling process, Multistep Template assisted soft lithography | 5mm/s | None | Micromixing, Pumping | Ref. 23 |
| 4 | 1,000 × 23,300 × 4,700 (height × length × width) | Fabricated with Low temperature co-fired ceramic tapes (LTCC) | Creeping flow | None | Micromixing | Ref. 24 |
| 5 | 225 × 11,700 (radius × length) | Single step Template assisted soft Lithography | Static | Thermal sensing and regulation, Optical sensing | Micromixing, Temperature dependent reactions | Current study |

## 4. Conclusion

A novel active micromixer device was successfully constructed using single-step template-assisted soft lithography. The active mixing volume was 1.86μl. The mixing in the device was achieved by Lorentz force acting on an enameled copper wire carrying AC signal in a static magnetic field. The temperature of the mixing chamber was actively maintained and the reaction occurring in the device was optically monitored using the fiber optical assembly. The repeatability in the ratio of harmonic frequency to fundamental frequency showed that the fabrication technique of the devices is standardized. The technique of integrating the principle of Lorentz force with single-step template-assisted soft lithography makes it a cost-effective and simple to fabricate method. Three different mixing reactions were performed to demonstrate the application of the device as having the potential to be a Lab-on-a-chip system.

Novel active micromixer platform based on Lorentz force for lab-on-a-chip application

Aniket Kandalkar [a], Nachiket Pathak [a, b], Atharva Kulkarni [c], Amit Morarka*[c]
[a] Institute of Bioinformatics and Biotechnology, Savitribai Phule Pune University, Ganeshkhind, Pune, Maharashtra, INDIA - 411007
[b] Indian Institute of Science, C.V. Raman Avenue, Bengaluru, Karnataka, INDIA-560012
[c] Department of Electronics and Instrumentation Science, Savitribai Phule Pune University, Ganeshkhind, Pune, Maharashtra, INDIA- 411007
*amitmorarka@gmail.com

# Supplementary Information Contents

**Supplementary Section 1**

**Fabrication of Micromixer device, its assembly, design, and the components used**

The mixing in the device was carried out by the D80 wire running along the length of the main channel. The total length of the copper wire used was 195mm out of which only 11.7mm of the wire was interacting with the sample in the main chamber. The device had a pocket of diameter 20mm and a height of 6mm at the bottom for the placement of a rare earth magnet. The device was mounted between the holders fixed onto a platform which was at a height of 0.125m. An in-house made electromagnet was mounted below the platform. The suspended weight was attached to the wire which was affixed to a load cell. The electromagnet provided an additional tension in the wire by attracting the suspended ferromagnetic weight. Thus, resulting in control over the tension in the wire by regulating the power of the electromagnet. The plastic guide provided directionality to the motion of the suspended weight. The suspended ferromagnetic weight was aligned exactly onto the electromagnet core as shown in Fig. S1. The electromagnet was made from an enameled copper wire of diameter 1mm with 500 turns wound over a plastic tube with an iron core inside it (Fig. S1).

The device was integrated with optical and thermal sensors. Optical fibers were integrated into the device for coupling the light through the main chamber. The Optical fibers were fixed in place through holes in the hollow cylindrical aluminium chassis. The main channel passed co-axially through the chassis. The optical path was fixed to 450µm. This provided efficient optical coupling of light from the source to the spectrophotometer through the main chamber. Hence, it assisted in real-time optical detection of the chemical reactions occurring inside the chamber. The optical fiber used in the device had a diameter of 2mm and a core of 1mm.

The heating element was electrically connected to the DC power supply. The heating of the chassis was monitored by the two PT-100. The PT-100 was mounted on the chassis by silver paste which provided thermal contact. The temperature sensors were electrically connected to the berg strip. The images of the same are shown in Fig. S2-S5**.**

The ends of the main channel were sealed using grease to avoid leaks as shown in Fig. S6. The grease was pumped manually inside the device through the two additional inlets and was completely isolated from the main chamber. The dimensions of the channels constructed in the device are provided in table S1.

The optical stroboscopic recording was done using a digital camera Canon Power shot SX730HS. The maximum amplitude at the resonant of the displaced wire was measured from the images. The sound of the oscillating wire was recorded by the microphone of the Samsung J2 mobile for the acoustic analysis of the resonant frequency of the oscillating wire. Resonant frequencies observed at three tensions applied on the wire were observed. The experiment was performed in triplicates with three different devices.

The reagents were pumped manually inside the device in a 1:1 volumetric ratio through the inlets using two Hamilton micro syringes (5µl and 10µl) for all the reactions performed in the device. Fig. S7 shows an assembled device and the complete setup of the device and additional peripherals.

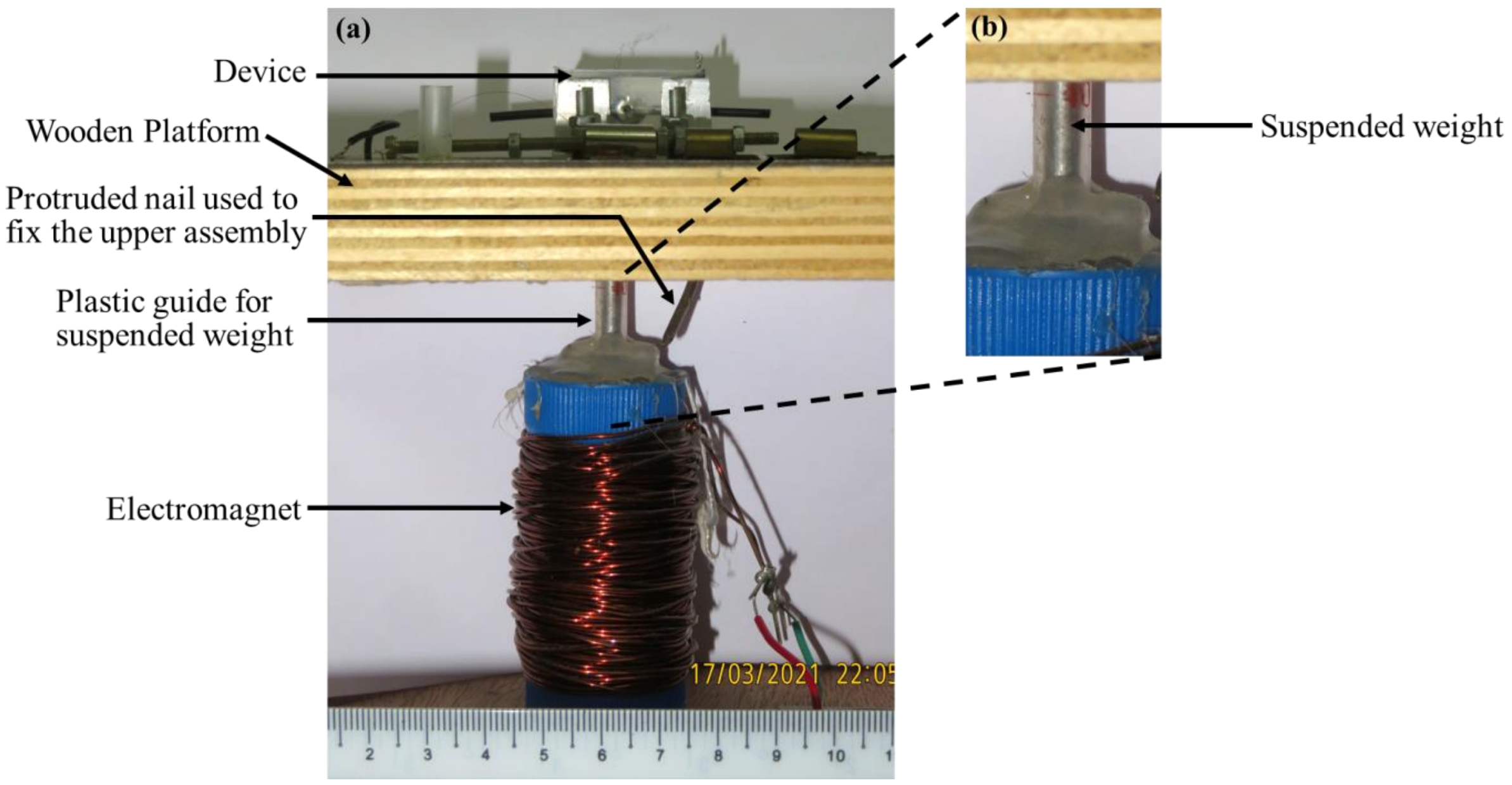

**Fig. S1**: Images of (a) Electromagnet placed below the suspended weight. (b) Inset representing the suspended weight

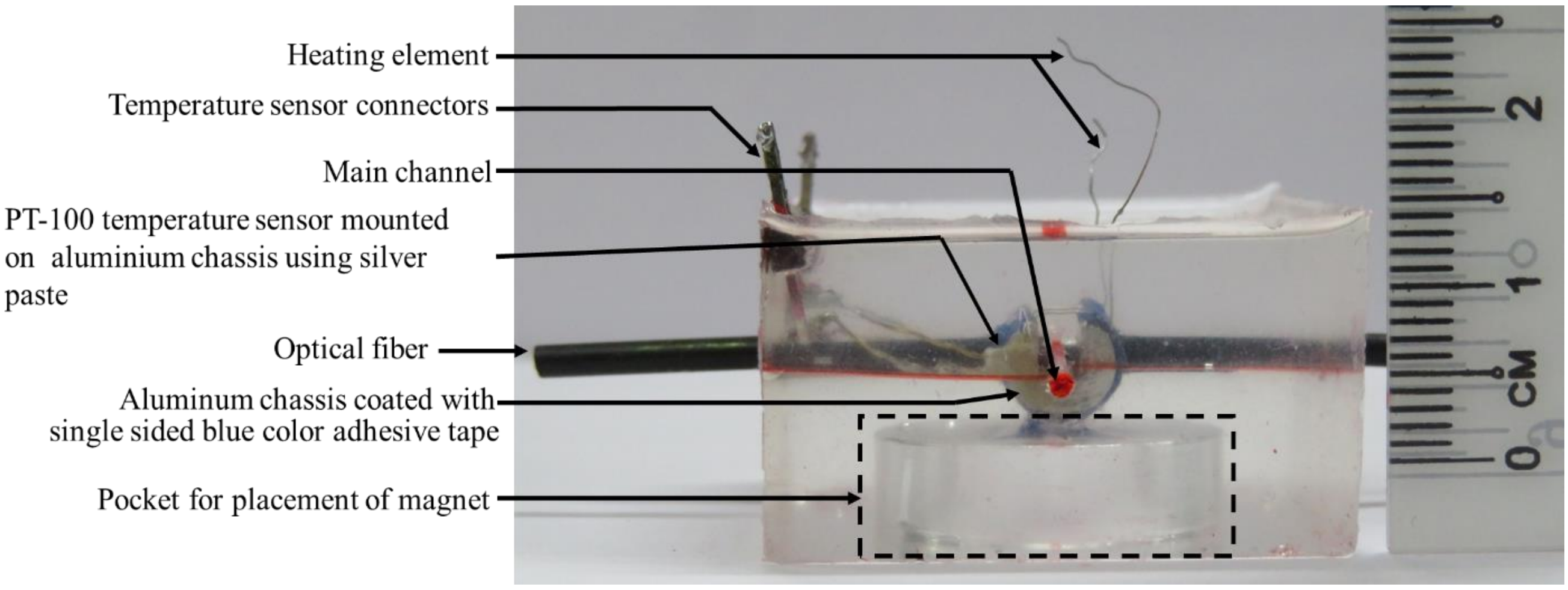

**Fig. S2**: Side view of the micromixer device from the main channel side. The figure shows the components embedded in the device.

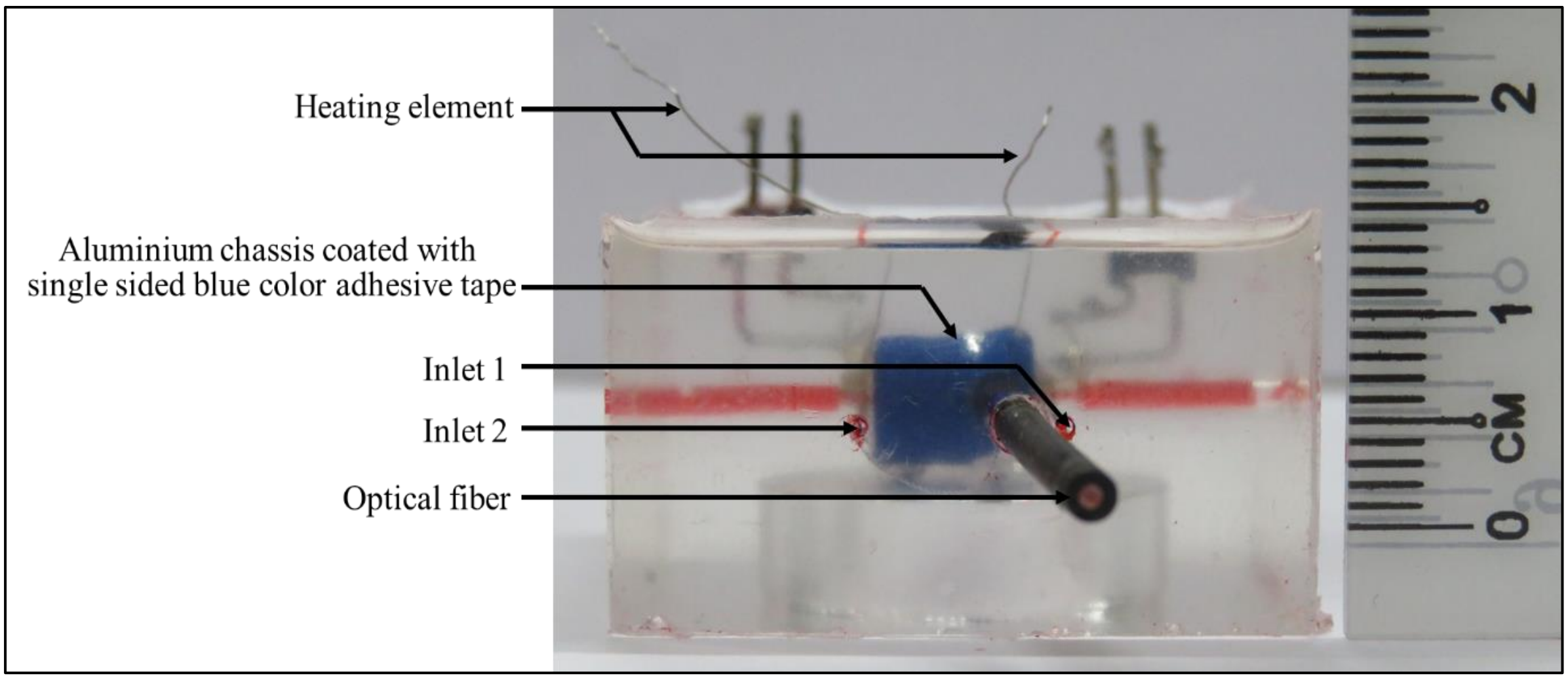

**Fig. S3**: Side view of the micromixer device from the inlet side. The figure shows the components embedded in the device.

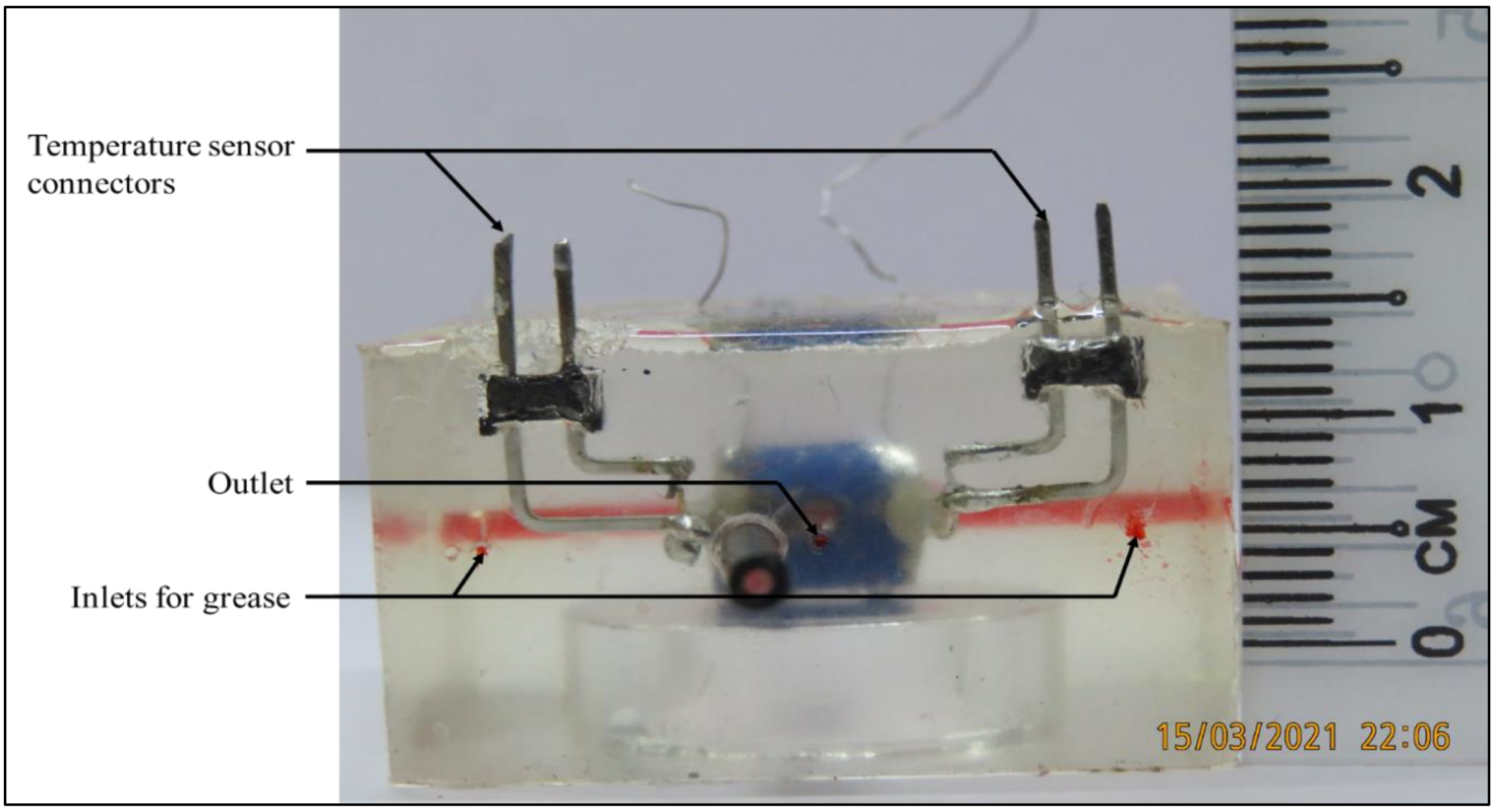

**Fig. S4**: Side view of the micromixer device from the outlet side. The figure shows the components embedded in the device.

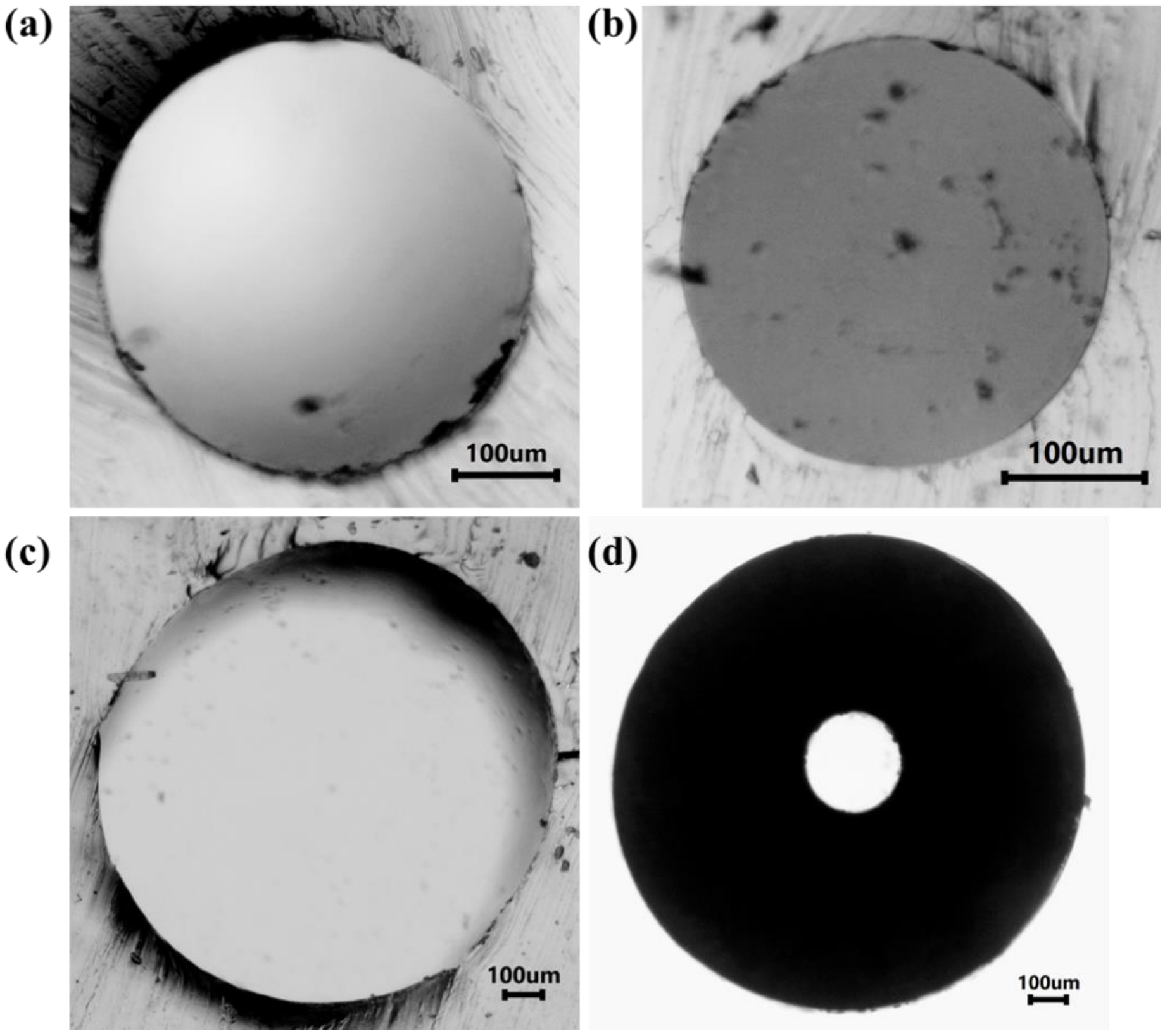

**Fig. S5**: Transverse section of (a) main chamber (b) connecting channel (inlets 1&2 and outlet) to the main chamber (c) ends of the main channel which holds bushing (d) bushing.

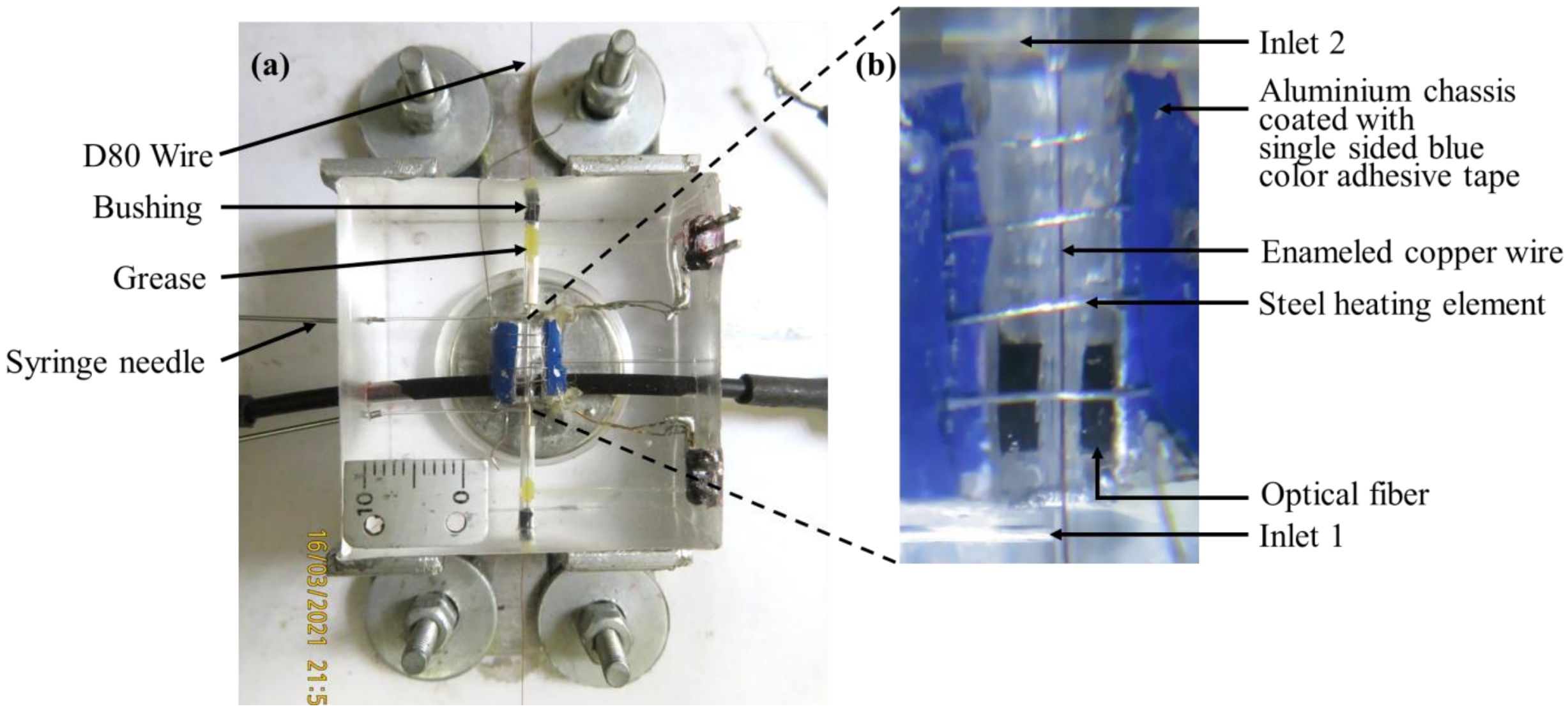

**Fig. S6**: Placement of the device and all attached peripherals. The wire was aligned in the device by the black colored bushings fixed at the ends of the main channel. (a) Microfluidic device integrated with the wire under tension and its ends sealed with grease. (b) Enlarged view of the main mixing chamber.

**Table S1**
Length of the channels in the device. The values represent the average dimensions of the three devices constructed.

| Sr. No. | Channels | Lengths ($10^{-2}$m) |
|---|---|---|
| 1 | Inlet 1 | 1.67 |
| 2 | Inlet 2 | 1.63 |
| 3 | Outlet | 1.80 |
| 4 | Main channel | 1.17 |

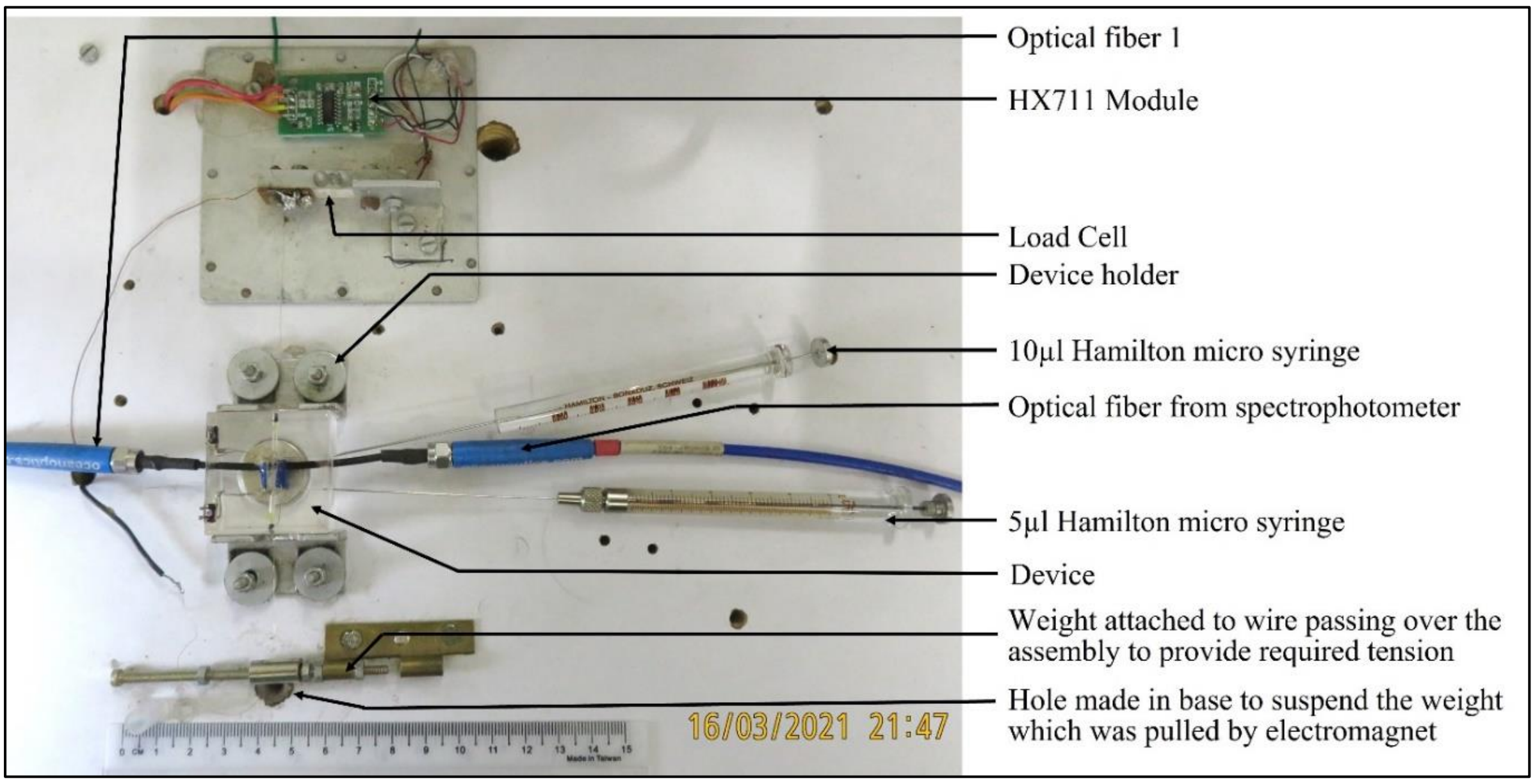

**Fig. S7**: Top view of the complete setup. The wire is connected to the load cell and the other end is passing over the pulley holding the suspended weight. A hole was made through the platform for the suspended weight to pass through it. The Optical fiber 1 was connected to the LED source used for optical detection. The light passing through the main chamber is coupled to the spectrophotometer as shown. The dimensions of the optical fiber integrated into the device were different from that of the fiber used to connect the spectrophotometer, hence an additional attachment was made to connect the two fibers. This assembly was optically coupled with the CCD-based spectrophotometer. Two Hamilton micro syringes were used for manually pumping the liquid in the device. The micromixer device was kept in place by the device holder.

## Supplementary Section 2

### Simulation result of the heating caused by D80

A thermal simulation of the current-carrying D80 was done in COMSOL. D80 of length 195mm was taken into consideration which passed through the micromixer system. The system was maintained in stagnant air and the external temperature was fixed to 26°C. The length of the wire outside the device was exposed to air. The normal current density passing through the wire was defined which caused ohmic heating. Heating caused by various currents and lengths of the wire was examined in the simulation (Table S2). An optimum value of current and length were selected. Mixing in the device was also possible at lower currents as it could cause less heating but at an expense of mixing time.

**Table S2:** Rise in temperature of D80 caused by ohmic heating with change in current passing through it.

| Sr. No. | Current (mA) | Temperature (°C) |
|---|---|---|
| 1 | 50 | 26.1 |
| 2 | 100 | 26.6 |
| 3 | 150 | 27.3 |
| 4 | 200 | 28.2 |
| 5 | 250 | 29.5 |
| 6 | 300 | 31.0 |

**Supplementary Section 3**

**Load cell calibration and study of the resonant frequency of the oscillating wire**

The data from the load cell was compared with the standard electronic weighing balance Shimadzu ATX124. Nine independent samples were used to compare load cell and the Shimadzu weighing balance readings. The plot represents the average of the three experiments (Fig. S9).

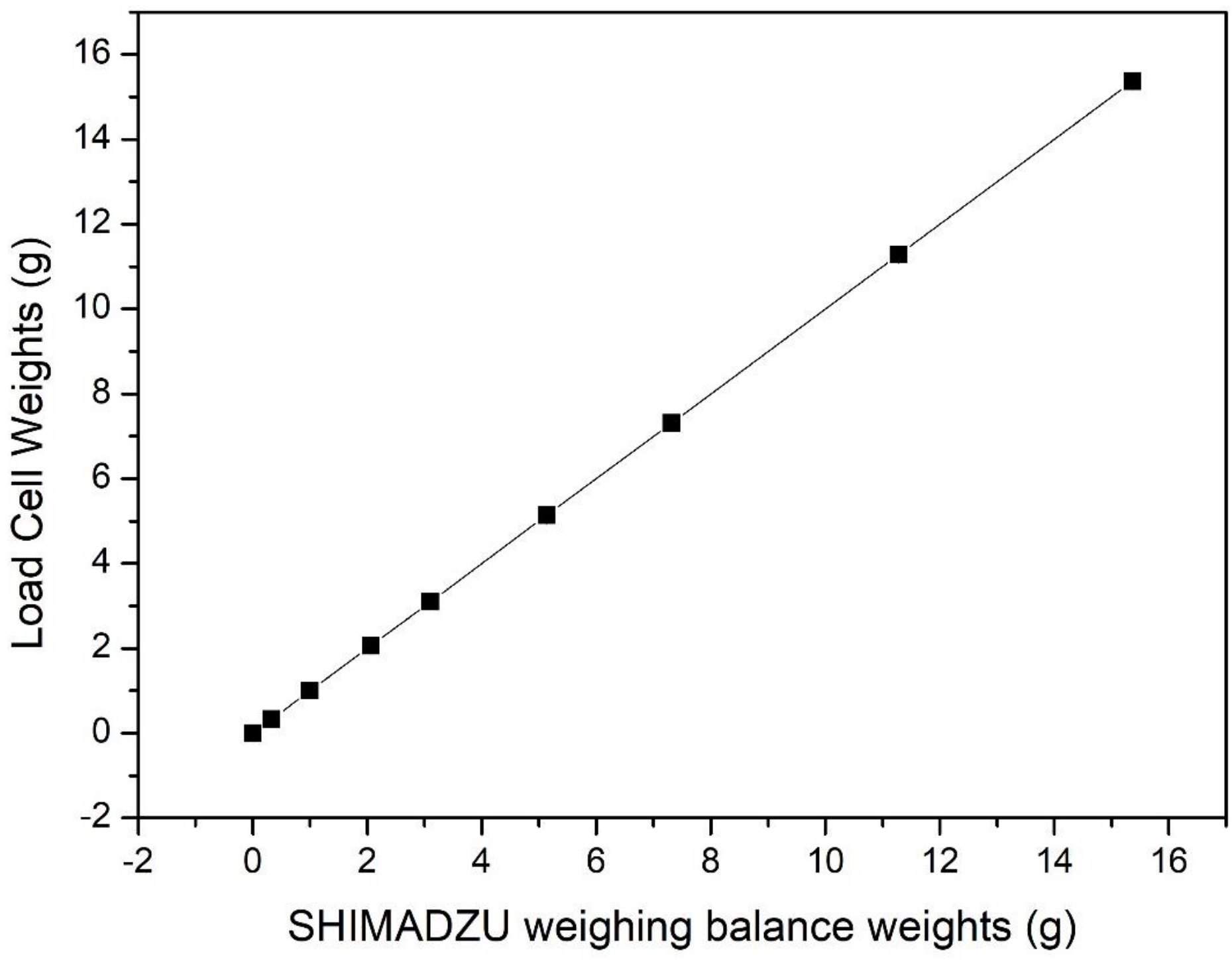

**Fig. S9:** Load cell calibration.

**Video recording of the oscillating wire**

Stroboscope video of the wire oscillations provided the resonant frequency of the wire. At the resonant frequency, the maximum amplitude of the wire oscillation was observed. The maximum displacement of the wire was measured from the still images captured from the video recordings. The amplitude measurement of the oscillating wire was done using ImageJ software. The distance was measured manually with respect to a fixed point in the image**.**

**Acoustic recording of the oscillating wire**

Measurement of the resonant frequency of the wire was also done by recording the acoustic of the wire oscillation. The acoustic measurement was done by placing the in-built microphone of the Samsung J2 mobile in contact with the side wall of the device. Further Fourier analysis of the audio recordings was done.

Complete data from video and audio recordings are shown in table S3.

**Table S3**

Harmonics of the wire observed at various tensions and their ratios

| Device | Tension (N) | Harmonics | Experiment 1 Frequency (Hz) | Experiment 2 Frequency (Hz) | Experiment 3 Frequency (Hz) | Average (Hz) | Ratio of harmonics to fundamental frequency |
|---|---|---|---|---|---|---|---|
| 1 | $39.2\times10^{-3}$ | Fundamental | 225 | 222 | 219 | 222 | - |
| | | $2^{nd}$ harmonic | 589 | 585 | 587 | 587 | 2.64 |
| | | $3^{rd}$ harmonic | 674 | 676 | 673 | 674 | 3.03 |
| | $68.6\times10^{-3}$ | Fundamental | 268 | 267 | 266 | 267 | - |
| | | $2^{nd}$ harmonic | 796 | 796 | 798 | 797 | 2.98 |
| | $107.8\times10^{-3}$ | Fundamental | 466 | 465 | 467 | 466 | - |
| | | $2^{nd}$ harmonic | 991 | 996 | 994 | 994 | 2.13 |
| 2 | $39.2\times10^{-3}$ | Fundamental | 206 | 207 | 206 | 206 | - |
| | | $2^{nd}$ harmonic | 523 | 529 | 524 | 525 | 2.54 |
| | | $3^{rd}$ harmonic | 610 | 612 | 612 | 611 | 2.96 |
| | $68.6\times10^{-3}$ | Fundamental | 272 | 277 | 274 | 274 | - |
| | | $2^{nd}$ harmonic | 834 | 846 | 837 | 839 | 3.06 |
| | $107.8\times10^{-3}$ | Fundamental | 313 | 320 | 326 | 320 | - |
| | | $2^{nd}$ harmonic | 979 | 976 | 973 | 976 | 3.05 |
| 3 | $39.2\times10^{-3}$ | Fundamental | 229 | 229 | 289 | 229 | - |
| | | $2^{nd}$ harmonic | 604 | 607 | 606 | 606 | 2.64 |
| | | $3^{rd}$ harmonic | 691 | 687 | 687 | 688 | 3.00 |
| | $68.6\times10^{-3}$ | Fundamental | 270 | 270 | 270 | 270 | - |
| | | $2^{nd}$ harmonic | 810 | 809 | 810 | 810 | 3.00 |
| | $107.8\times10^{-3}$ | Fundamental | 337 | 338 | 336 | 337 | - |
| | | $2^{nd}$ harmonic | 991 | 989 | 991 | 990 | 2.93 |

## Supplementary Section 4

### Thermal simulation and experimentation of the device and calibration of the thermocouple

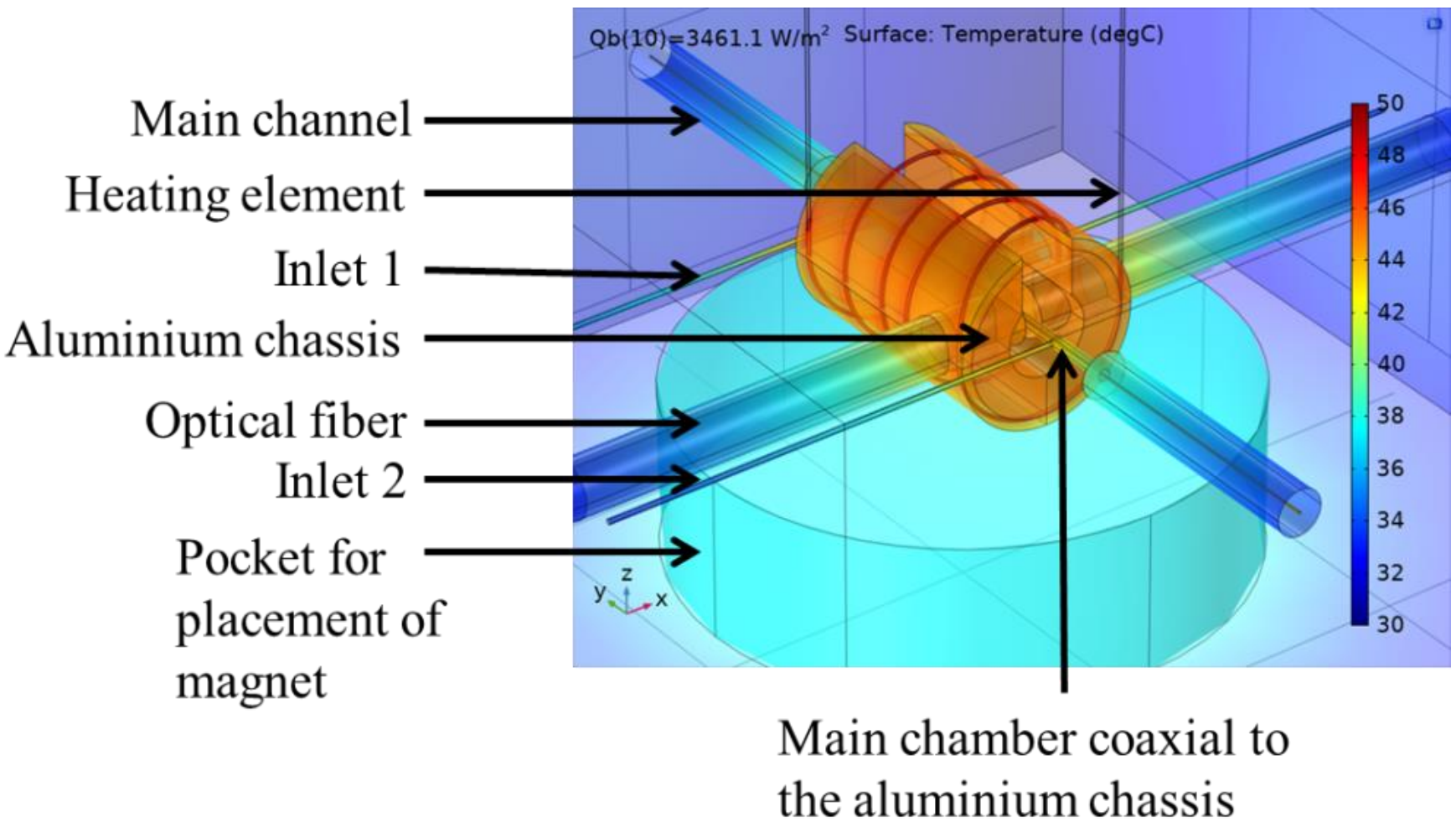

**Fig. S10(a):** Enlarged 3D prospective view from thermal simulation of the micromixer highlighting the device parts.

Thermal simulation results of micromixer using COMSOL 5.5. The simulations show that there is a uniform distribution of temperature in the device for a temperature range of 25-74°C. The top view of the device is plotted in the following two-dimensional graphs.

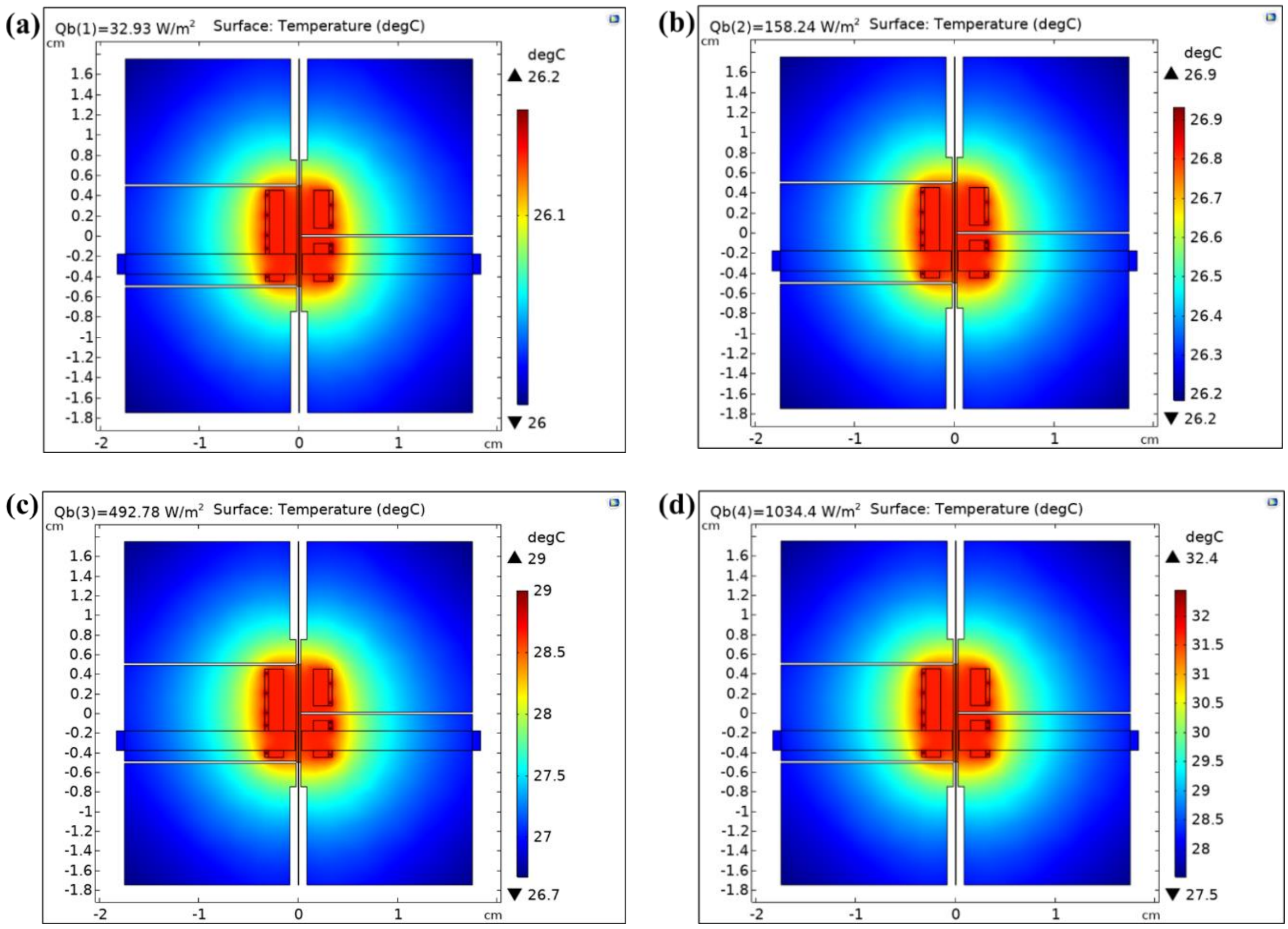

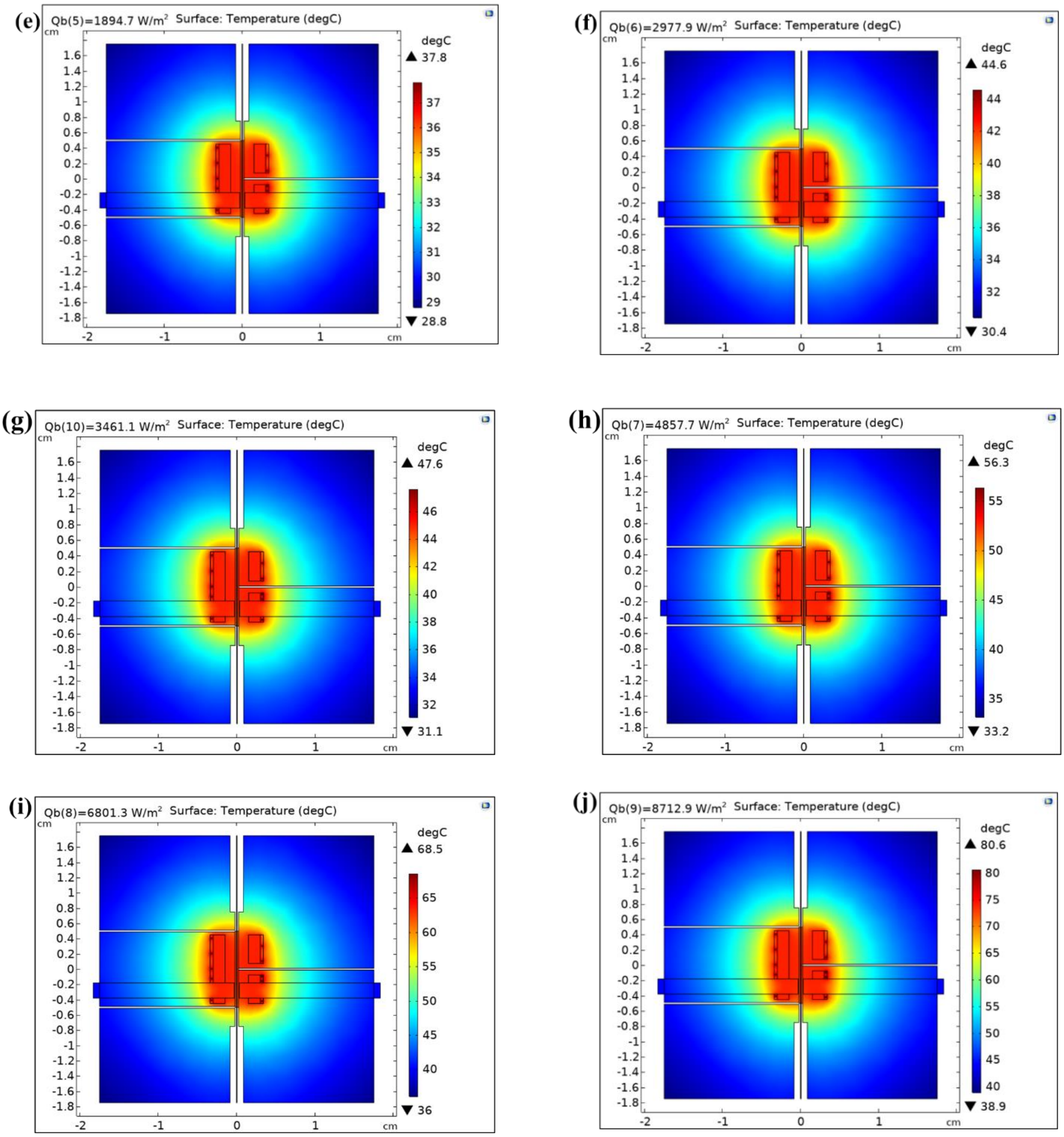

**Fig. S10**: Longitudinal section of device passing through center showing thermal simulation results of the micromixer device. The main channel temperature as shown in figure: (a) = 26.13°C, (b) = 26.81°C, (c) = 28.65°C, (d) = 31.69°C, (e) = 36.44°C, (f) = 42.42°C, (g) = 45.09°C, (h) = 52.80°C, (i) = 63.54°C, (j) = 74.18°C.

## Calibration of the thermocouple

The K-type thermocouple was made manually by wielding the two metals wire of diameter 200μm (Chromel and Alumel) using a Butane gas torch. The wielded spot was spherical which had a size comparable to the diameter of the main channel. The calibration of the thermocouple was done by placing it in between two Peltier devices. The Peltier devices were used to maintain the constant temperatures. A reference PT-100 was placed with a thermocouple for calibration. The data obtained from the thermocouple is compared with the commercially available PT-100. The readings were taken in triplicates, no hysteresis was observed while sensing the temperature. The in-house made K-type thermocouple was read using MAX6675 module using Arduino UNO with a resolution of 0.25°C. Their calibration charts with respect to the commercially available PT-100 are shown in Fig. S11. The temperature calibration was done for a range from 24°C to 86°C for the thermocouple. For all the further thermal studies of devices, thermocouple 1 was used.

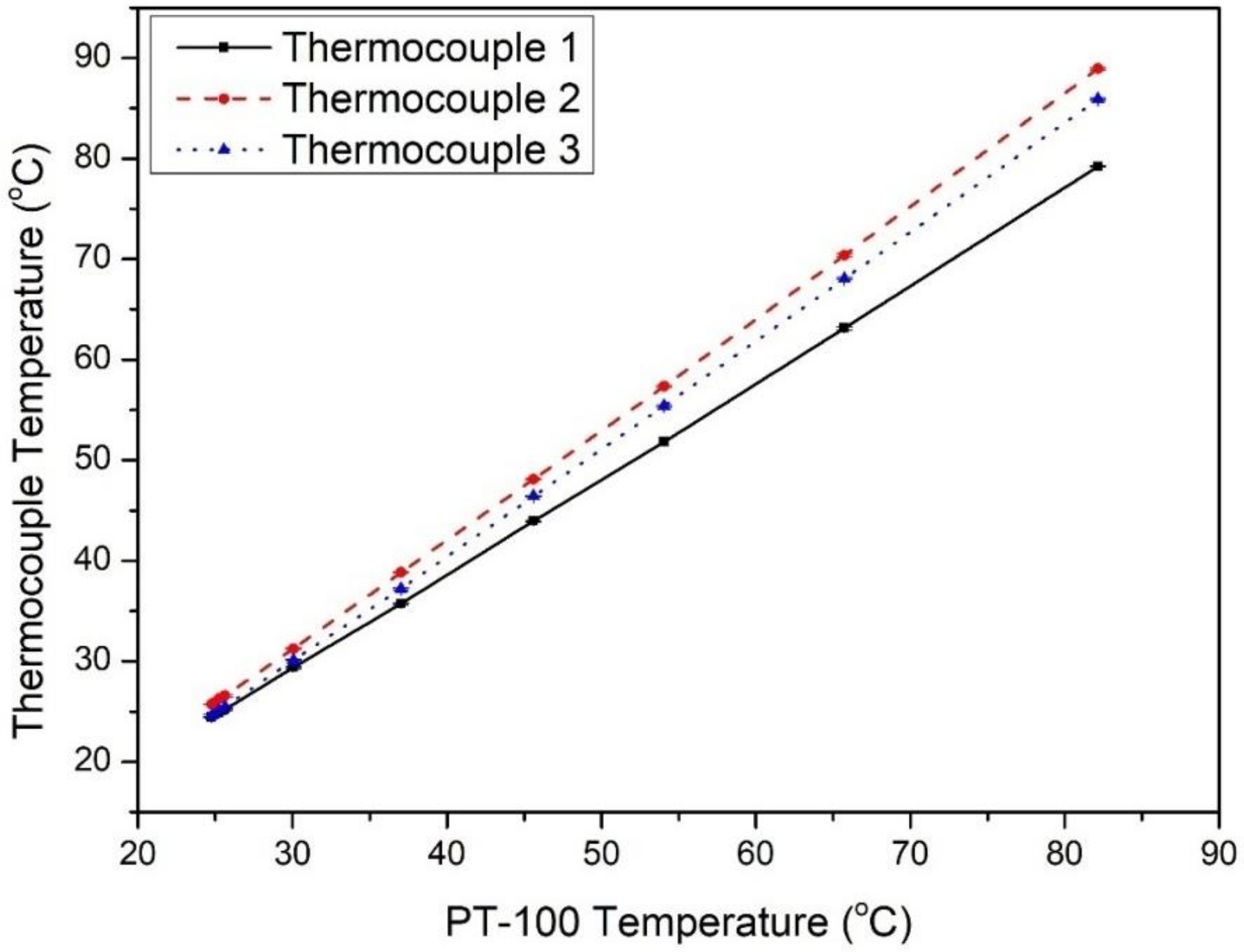

**Fig. S11**: Thermal calibration of the in-house made thermocouple.

## Thermal calibration of three micromixer devices

Thermal calibration of three independent devices was done using thermocouple 1. The thermocouple and PT-100 readings were taken when the system reached thermal equilibrium. All the readings were taken at room temperature. PT-100 1 and PT-100 2 represent the sensor attached to the ends of the aluminum chassis. Fig. S12 shows a plot of the temperature calibration of the chassis and the temperature of the main chamber for three devices. To prove the repeatability of the device fabrication, three devices were made and characterized.

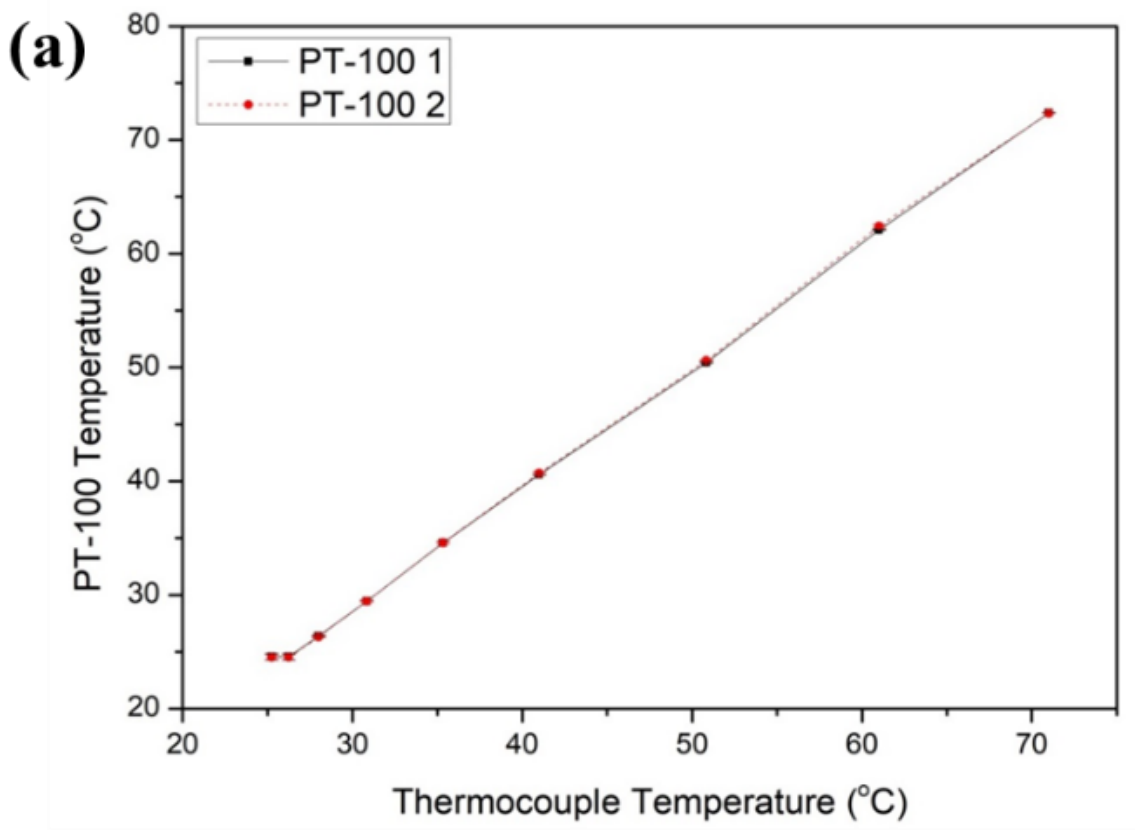

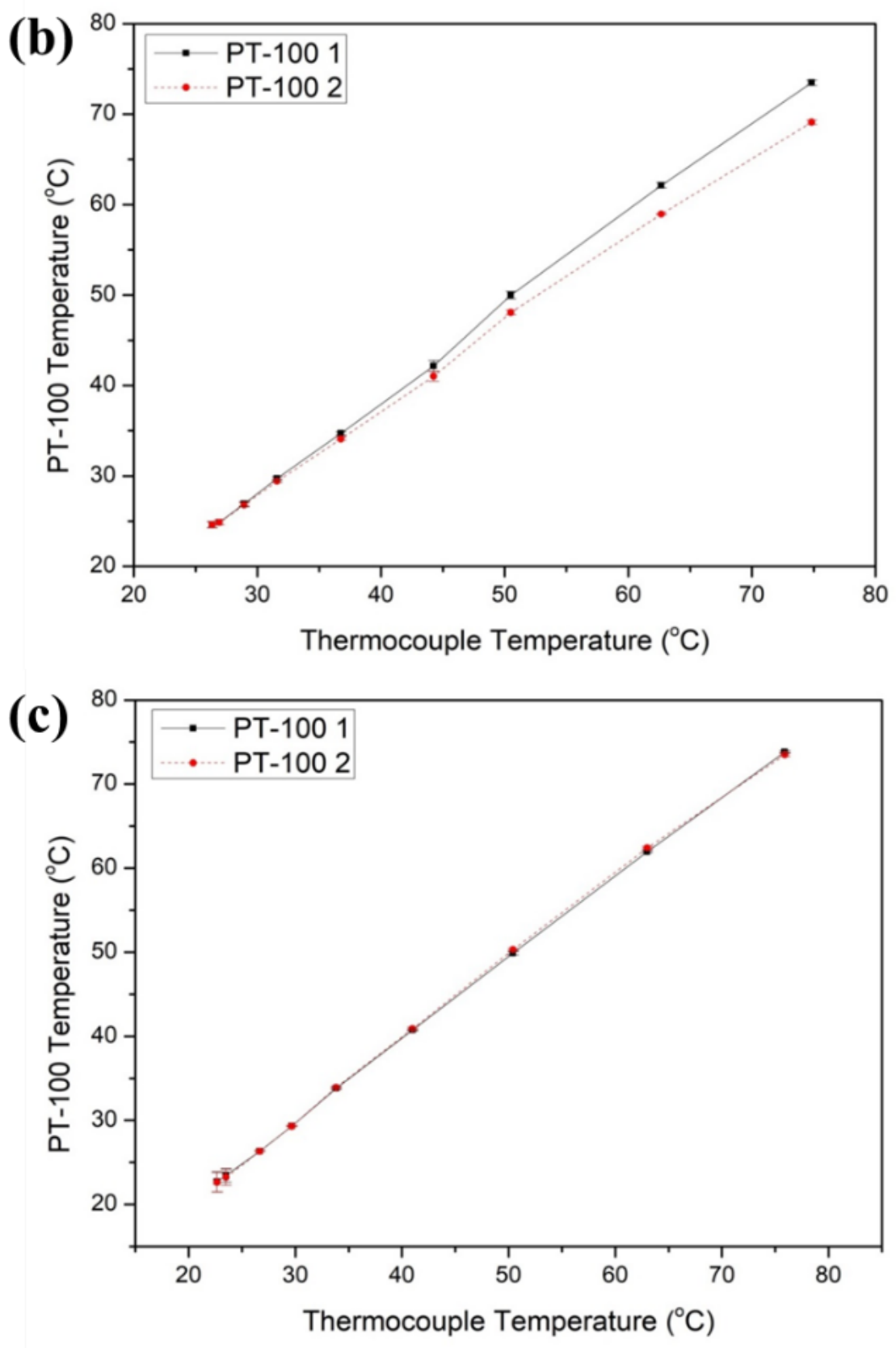

**Fig. S12**: Thermal calibration of the devices. The plot compares temperature reading from the two PT-100 with that of the thermocouple. (a) Device 1, (b) Device 2, (c) Device 3

**Supplementary Section 5**

**Optical calibration of the Micromixer device**

Calibration of the optical setup integrated into the device was done by measuring the intensity counts of the Potassium permanganate ($KMnO_4$) solution. Water was taken as a control for the experiment. A stock solution of 0.01M $KMnO_4$ was prepared. Further eleven serial dilutions of 0.5 times were made. The results show that the optical assembly could couple light efficiently and aid the spectrophotometer to detect the dilutions accurately. The solution was pumped into the device through either of the inlets. Average relative Intensity counts obtained from the CCD-based Ocean Optic HR4000 were reported from the three experiments (Fig. S13**).** The optical assembly could measure concentration accurately up to 0.156mM of $KMnO_4$.

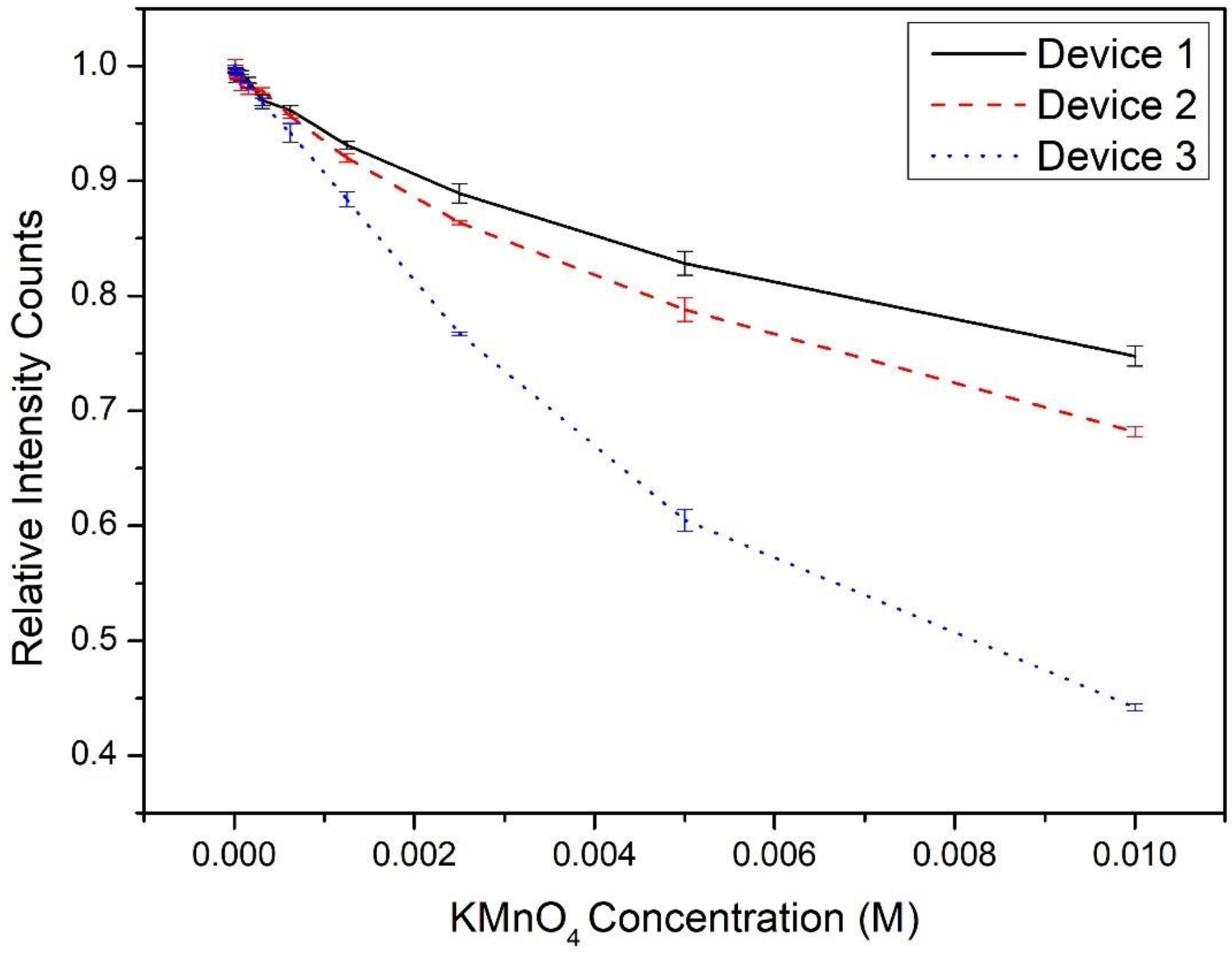

**Fig. S13**: Optical calibration of the 3 devices.

**Supplementary Section 6**

**Images of mixing in the Micromixer device**

**Images of ink and water mixing carried out by diffusion in the micromixer device**

Images were captured from the video recordings of the mixing performed in the device. The captured images were cropped. The contrast and the brightness of the images were changed for better visibility.

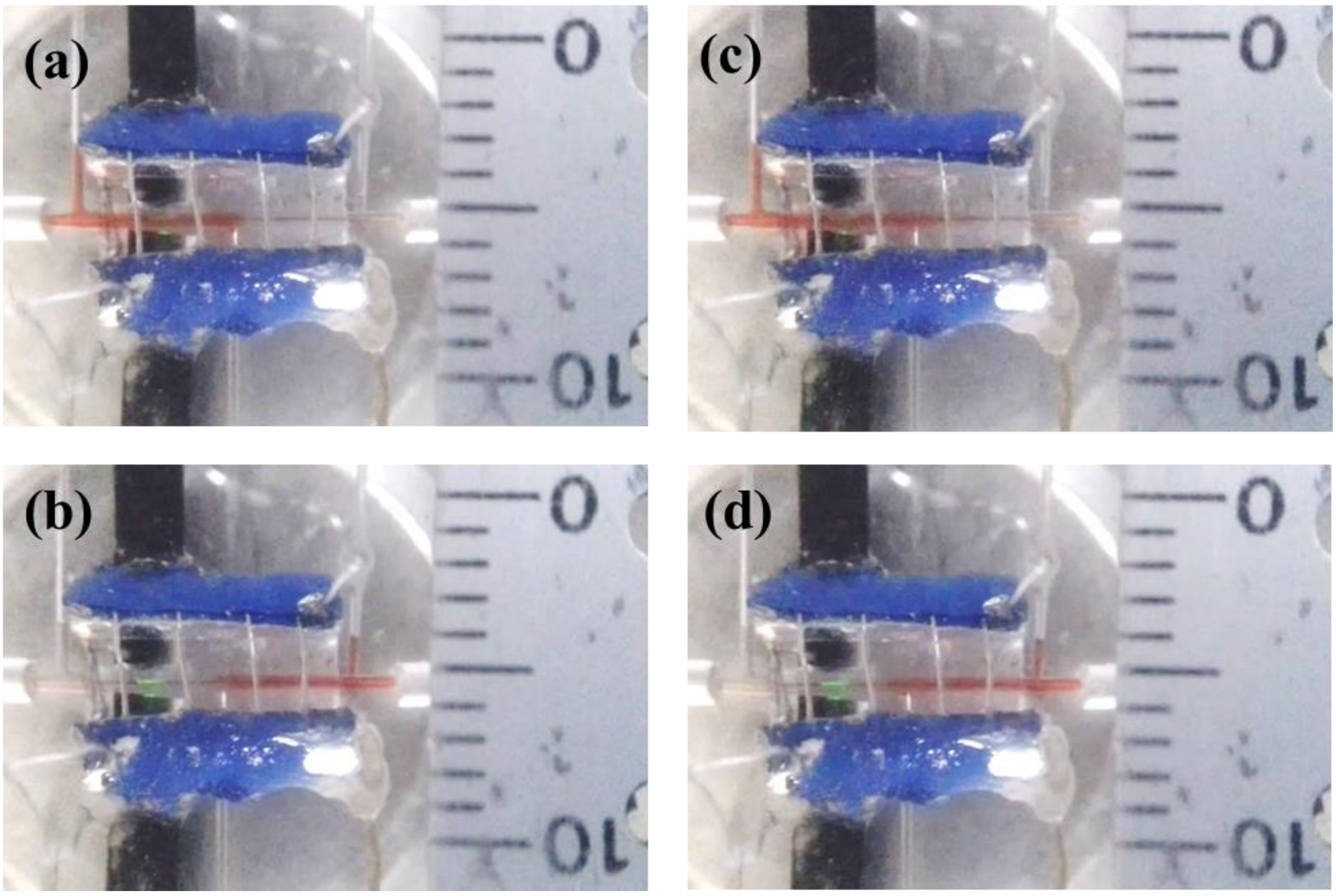

**Fig. S14:** The images of mixing of ink and water in the micromixer. Images of the mixing that were carried out by diffusion. [(a), (b)] Image of the mixing by diffusion taken at time 0min. Where (a) represent ink on the side of the sensor, and (b) represents water on the side of the sensor. The solution (a) after 10min is shown in (c) as the partially mixed solution. Similarly, the solution (b) after 10min is shown in (d).

**Images of active mixing in the device**

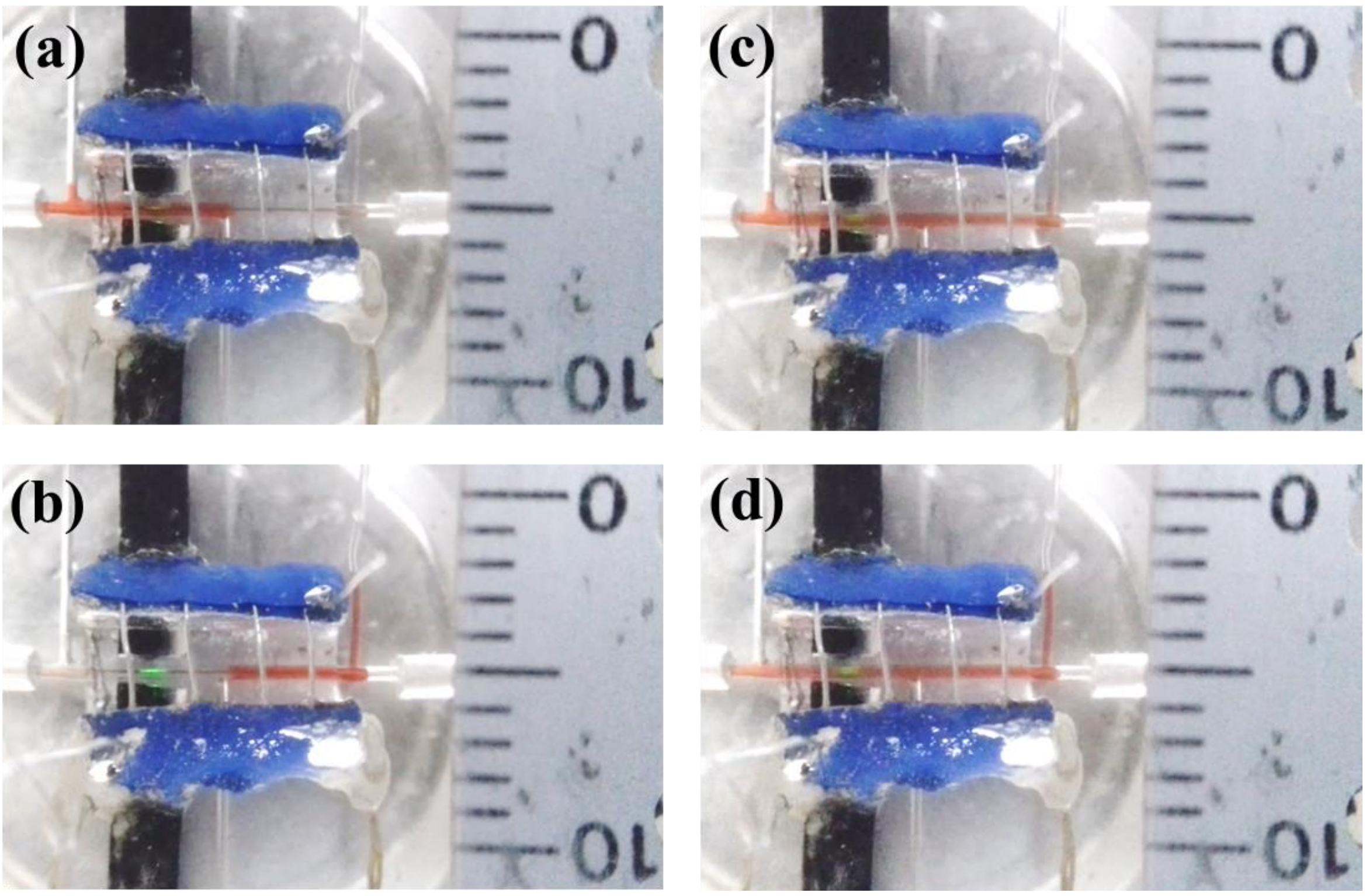

**Fig. S15**: Images of the active mixing of ink and water carried out in the device. [(a), (b)] Images of active mixing in the device at time 0min. Where (a) represent ink on the side of the sensor, and (b) represents water on the side of the sensor. (c), (d) represents mixed solutions after 10min with respect to time 0min in (a), (b) respectively.

**The images of temperature dependent reaction in the micromixer**

Images were captured from the video recordings of the chemical reaction performed in the device. The captured images were cropped. The contrast and the brightness of the images were changed for better visibility.

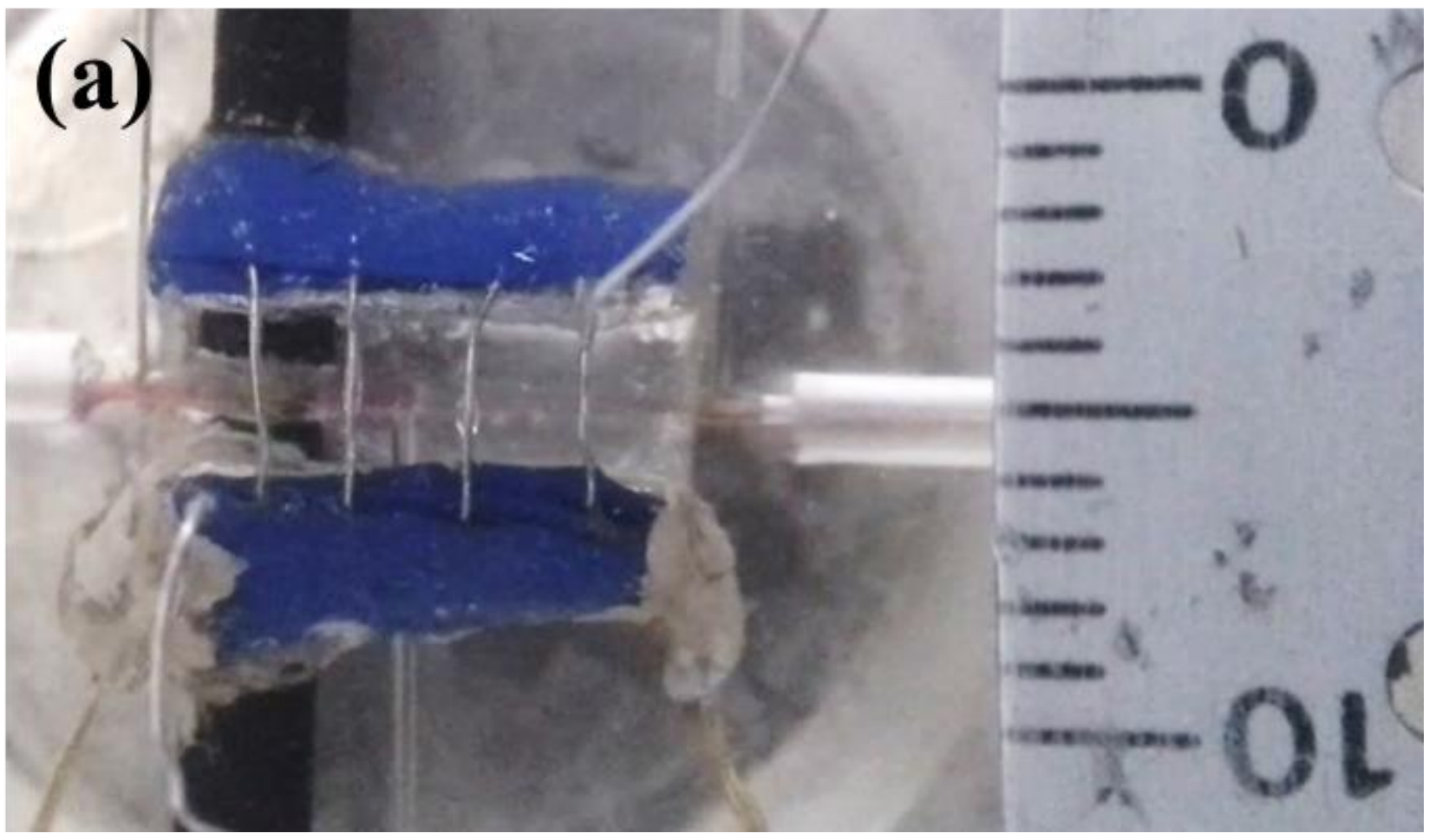

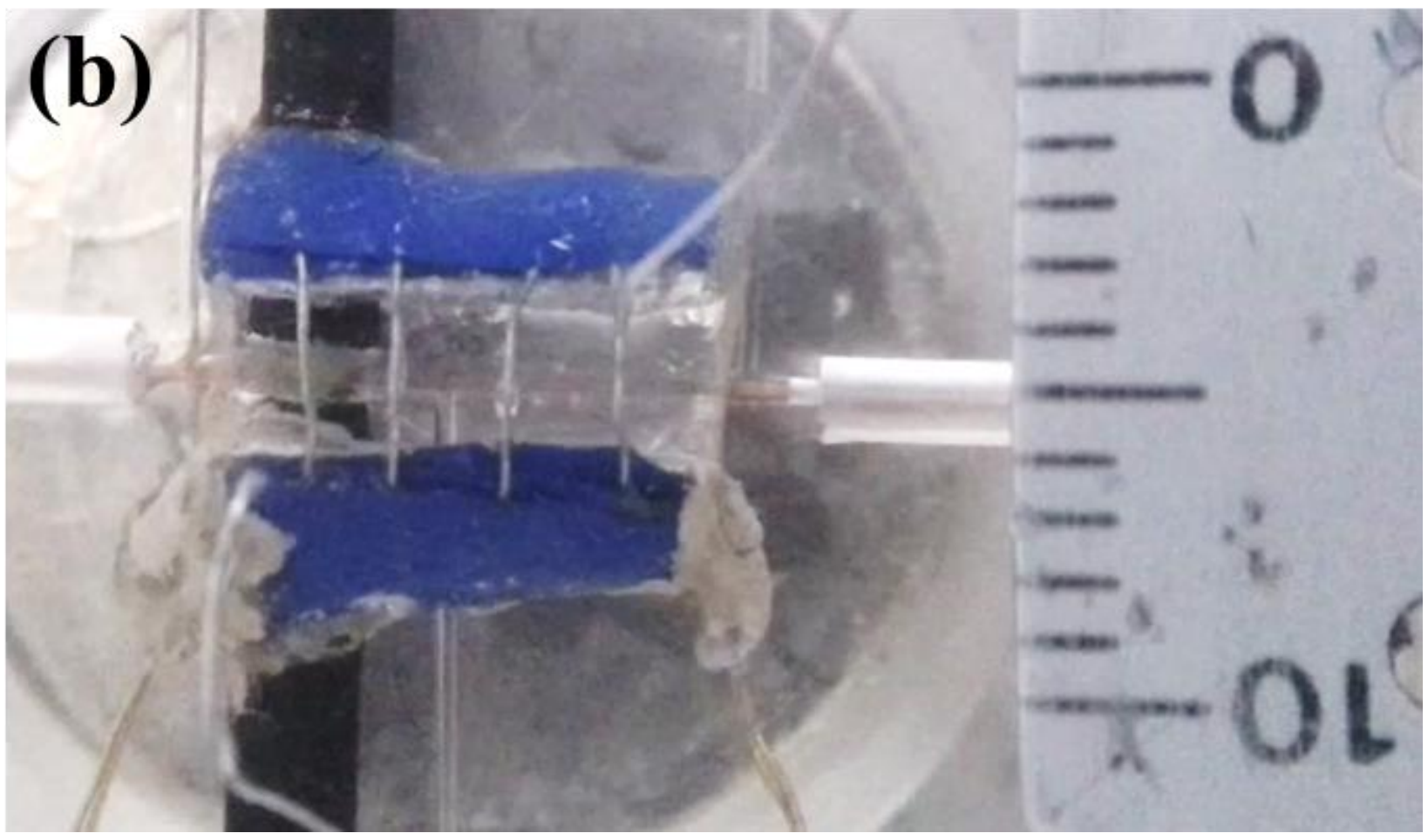

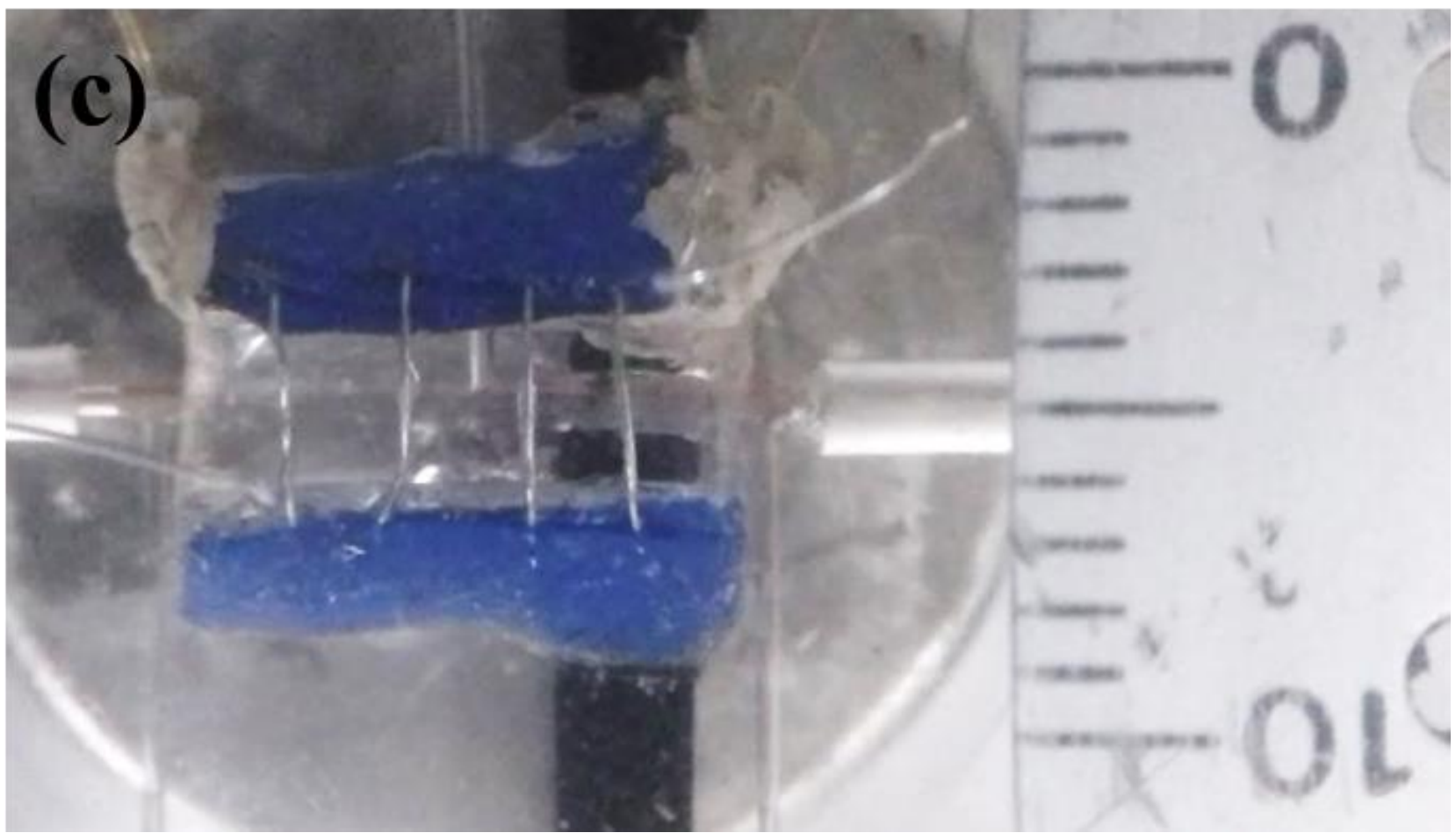

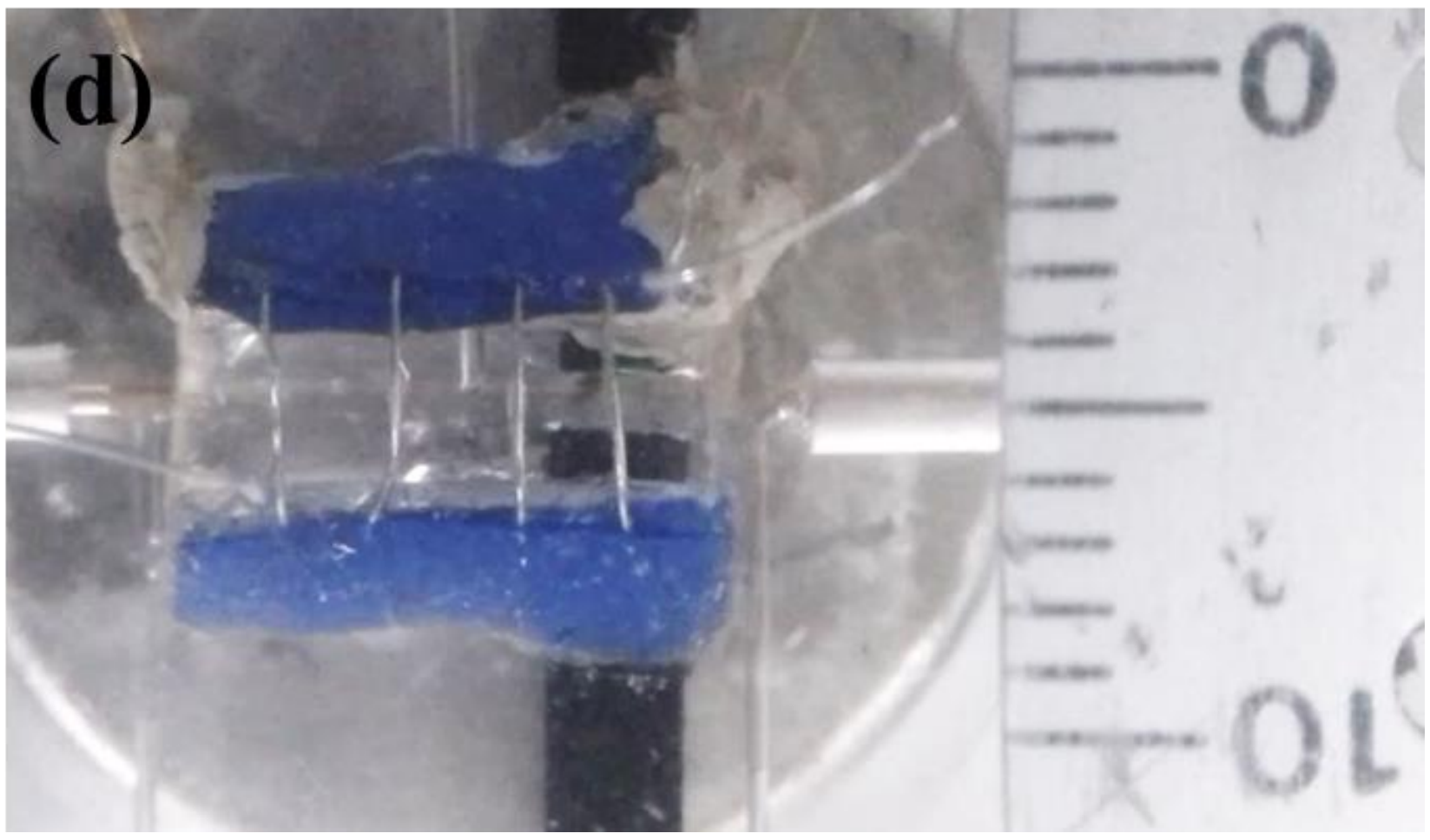

**Fig. S16**: The images of temperature-dependent reaction in the micromixer. In all these reactions active mixing was carried out at a specific temperature. (a), (b) are the images of the reaction carried out at 20°C and (c), (d) carried out at 45°C. At time 0min (a), (c) a little shade of pinkish color can be seen in the MC near the fiber optic sensor. After time 570s a gradient of pinkish shade can be seen in (b) whereas a completely transparent solution is observed in (d).